\documentclass[aps,prb,twocolumn,floatfix,longbibliography,nofootinbib,superscriptaddress]{revtex4-2}
\usepackage[utf8]{inputenc}
\usepackage{natbib}
\usepackage{graphicx}
\usepackage{xcolor}
\usepackage[tbtags]{amsmath}
\usepackage[colorlinks, linkcolor=blue]{hyperref}
\hypersetup{colorlinks,allcolors=blue}
\usepackage{amssymb}
\usepackage{amsmath}
\usepackage{gensymb}
\usepackage{float}
\usepackage{comment}
\usepackage{amsmath}
\usepackage{tabularx,graphicx}
\usepackage{epstopdf}
\usepackage{latexsym}
\usepackage{color, colortbl}
\usepackage{psfrag}
\usepackage{bm}
\usepackage{lmodern}
\usepackage{fix-cm}
\usepackage{xfrac}
\usepackage{color}    
\usepackage{lipsum}
\usepackage{textcomp}
\usepackage{amsmath}
\usepackage{mathtools}
\usepackage{subfigure}
\usepackage{titlesec}
\usepackage{slashed}
\usepackage{multirow}
\usepackage[normalem]{ulem}
\usepackage{svg}
\renewcommand{\vec}[1]{\boldsymbol{#1}}

\newcommand{\bcen}{\begin{center}}
\newcommand{\ecen}{\end{center}}
\newcommand{\btab}{\begin{tabular}}
\newcommand{\etab}{\end{tabular}}
\newcommand{\bdes}{\begin{description}}
\newcommand{\edes}{\end{description}}

\newcommand{\ul}{\underline}
\newcommand{\beq}{\begin{equation}}
\newcommand{\eeq}{\end{equation}}
\newcommand{\bea}{\begin{eqnarray}}
\newcommand{\eea}{\end{eqnarray}}

\newcommand{\bary}{\begin{array}}
\newcommand{\eary}{\end{array}}
\newcommand{\benum}{\begin{enumerate}}
\newcommand{\eenum}{\end{enumerate}}
\newcommand{\bitem}{\begin{itemize}}
\newcommand{\eitem}{\end{itemize}}

\renewcommand{\vec}[1]{\boldsymbol{#1}}
\newcommand{\eqn}[1] {eqn.~(\ref{#1})}

\newcommand{\Fig}[1]{Fig.~\ref{#1}}

\makeatletter

\newcommand{\Rmnum}[1]{\expandafter\@slowromancap\romannumeral #1@}
\makeatother

\begin{document}

\title{Tunable Topological Phases in Quantum Kirigamis}

\author{Rahul Singh}
\email{rahulsingh21@iitk.ac.in}
\affiliation{Department of Physics, Indian Institute of Technology Kanpur, Kalyanpur, UP 208016, India}

\author{Adhip Agarwala}
\email{adhip@iitk.ac.in}
\affiliation{Department of Physics, Indian Institute of Technology Kanpur, Kalyanpur, UP 208016, India}

\begin{abstract}
Advances in engineering mesoscopic quantum devices have led to new material platforms where electronic transport can be achieved on foldable structures. 
In this respect, we study quantum phases and their transitions on a Kirigami structure, a Japanese craft form, where its individual building blocks are topologically non-trivial. 
In particular, we find that by mechanically deforming the Kirigami structure one can engineer topological phase transitions in the system. Using a multi-pronged approach, we show that the physics of the system can be captured within a transfer matrix formalism akin to Chalker-Coddington networks, where the junctions describe scattering between effective non-Hermitian one-dimensional edge channels. We further show that the nature of the Kirigami structure can affect the critical folding angles where the topological phase transitions occur. Our study shows the rich interplay between topological quantum phenomena and structural configuration of an underlying Kirigami network reflecting its potential as an intriguing platform. 
\end{abstract}

\maketitle

\section{Introduction}

The possibility of mechanically deforming mesoscopic quantum devices holds promise for the next generation of quantum technologies \cite{rogers2010materials, santangelo2023making, el2022mechanical, el2021digital,zhai2021mechanical, chen2020kirigami, blees2015graphene, KirigamiPRA2023, liu2018nano, LishuaiADvMat2023} and raises interesting theoretical questions \cite{Castro_PRL_2018, Flachi_PRD_2019, Imaki_PRD_2019}. The experimental demonstration of graphene Kirigamis \cite{blees2015graphene}, their bending rules \cite{GrossoPRL2015}, and high throughput computational search \cite{Hanakata_PRL_2018, Campbell_kirigami_2016, Campbell_kirigami_2014} have established them as a unique platform. 
While a lot of work has gone into engineering the mechanical properties of such systems \cite{Liu_PRR_2022, LishuaiADvMat2023}, little effort has been given to uncover the transport properties of such systems \cite{Campbell_kirigami_2016, Chen2016Topological,mortazavi2017thermal, zheng2022continuum}.

In this work, we pose the question, can mechanical modulation engineer a quantum phase transition in such Kirigami-based systems? In particular, in recent years, topological phases of matter have ushered in a new paradigm where materials can host robust quantized transport on the edges even while the bulk remains insulating \cite{Ludwig_PS_2015, Chiu_RMP_2016, Hasan_RMP_2010, Qi_RMP_2011, asboth2016short}. We explore whether Kirigami networks built out of such topological materials can show phase transitions as a function of the mechanical deformation. We establish theoretically that indeed such phase transitions are realizable in these systems. Using a multi-scale approach, starting from minimal models, effective junction modelling, and scattering-matrix calculations we show that such systems behave as `tunable' Chalker-Coddington networks \cite{chalker1988percolation, Chalker1996networkmodels,kramer2005random, Lee_PRL_1993, Snyman_PRB_2008, Cho_PRB_1997}, usually studied for disordered topological systems \cite{GruzbergPRBkagome, ChalkerPRL2020Floquet, GruzbergPRLFinite-Size, PhysRevX2015Measurement, Chalker2001Thermalnetwork, Pasek2014Networkmodels, Bilayergraphenenetwork2021} where the deformation of the network can be included to determine the junction parameters of the network.

\begin{figure}
    \centering
 \includegraphics[width=0.9\columnwidth]{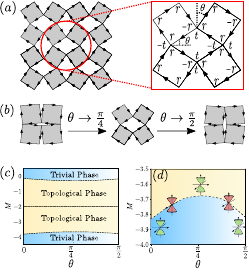}
\caption{{\bf Model and Phase Diagram:~}(a) A Kirigami network made up of square blocks where each block is a Chern insulator hosting edge states. The zoomed region shows the effective unit cell and scattering matrix elements $r$ and $t$. 2$\theta$ is the inter-block angle. 
(b) The network as a function of $\theta$ for $\theta \sim 0, \frac{\pi}{4}$ and $\frac{\pi}{2}$. (c) Numerically obtained phase diagram as a function of $\theta$ and $M$. 
(d) Zoomed region of (c) near $M \sim -3.8$. The critical dashed line corresponds to $r(\theta)=t(\theta)$. The effective dispersion of the network undergoes a  Dirac cone closing such that the blue (yellow) corresponds to the trivial (topological) phase.}
\label{fig:squarekiri}
\end{figure}

We model a foldable Kirigami as a network of two-dimensional lattice blocks attached at the vertices (see \Fig{fig:squarekiri}(a)). If each of the block is topological such that it hosts a chiral edge state, then the vertices can be modeled by a scattering matrix characterized by reflection and transmission amplitudes $r$ and $t$ (see the zoomed region in \Fig{fig:squarekiri}(a)). Interestingly as the structure unfolds, the angle $2\theta$ between the blocks can be tuned as shown in \Fig{fig:squarekiri}(b). In the regime where a single block becomes topological as a function of a microscopic parameter $M$ (say), we show that $\theta$ can be used to engineer a topological phase transition where the effective Hamiltonian for the {\it network} shows Dirac cone closings as shown schematically in \Fig{fig:squarekiri}(d). This renders the network trivial even when each of the blocks are themselves topological. The complete phase diagram is shown in \Fig{fig:squarekiri}(c).

{{In section~\ref{sec:model}  and~\ref{sec:kirigamimorpho} we discuss the microscopic model and the physical properties of the Kirigami. In section~\ref{sec:method} we discuss the utility of the three-step methodology which we employ. In section~\ref{sec:results} we present all the results and discuss the phase diagram of the system. In section~\ref{sec:triang} we study another triangular Kirigami network using the same formalism. In section~\ref{sec:discussion} we conclude our work. Additional numerical results and parameter dependencies have been relegated to Appendix~\ref{sec:appen}. }}

\begin{figure*}
    \centering
\includegraphics[width=1.8\columnwidth]{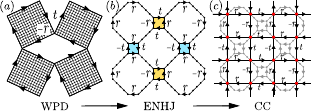}
\caption{{\bf Multiscale approach:~} The problem is analyzed within a three-stage process where in the first step (a) wave-packet dynamics (WPD) is used to characterize a single node of junctions between Kirigami blocks (see section~\ref{sec:Wave-packet-dynamics}). (b) An effective non-Hermitian junction (ENHJ) to model the complete network with two types of nodes $N_1$ and $N_2$ as shown with yellow and blue colors respectively (see section~\ref{sec:ENHJ}). (c) A Chalker-Coddington network where a median lattice is used to understand the topology of the full network (see section~\ref{sec:CC}).}
\label{fig:method}
\end{figure*}
\section{Model}
\label{sec:model}

We model the Kirigami with a periodic arrangement of blocks with associated nodes governed by scattering matrices. We consider the microscopic Hamiltonian governing each of the blocks to be the paradigmatic Bernevig-Hughes-Zhang (BHZ) model \cite{bernevig2006quantum}. The Hamiltonian in momentum space is given by 
\beq
H_{k_x,k_y} =\sin(k_x) \sigma_x  +  \sin(k_y) 
\sigma_y + (M+2-\cos k_x-\cos k_y)\sigma_z 
\label{BHZham}
\eeq
where $M$ is the parameter that tunes topological and trivial phases such that for $M>0$ and $M<-4$ we have a trivial phase, while for $0>M>-2$ and $-2>M>-4$ we have Chern insulating phases with Chern number 1 and -1 respectively. All parameters are measured in terms of the bare hopping scale which is set to unity. We assume that every block is thermodynamically large while the lattice constant $a =1$ and is much smaller than the block size $L^2$. In the topological phase, the edge retains chiral edge states while the bulk states remain gapped. In particular Chern number $C=-1$ ($C=1$) phase leads to clockwise (anticlockwise) edge states. { The whole network containing $L_B \times L_B$ number of Kirigami blocks can thus be represented by a tight-binding Hamiltonian with a total size of single-particle Hilbert space dimension $N_T= 2 L^2 L^2_B$ since every site contains two orbitals $A$ and $B$. To understand the transport property of such a network we use a multi-scale methodology with three different stages applied to the morphology of the Kirigami structure. We first briefly discuss the Kirigamis we intend to study before detailing its transport properties.

\section{Kirigami morphology}
\label{sec:kirigamimorpho}

Before we specifically comment on methods to calculate transport quantities it is useful to discuss the mechanical properties of Kirigamis and how they behave under deformation thus affecting the underlying microscopic Hamiltonian. Kirigamis, particularly of experimental and technological interests can either be parallel cut or cross cut \cite{tao2023Engineering, rafsanjani2017buckling, lee2020multiaxially}. The parallel-cut version of Kirigami drew a lot of theoretical interest due to its realization in graphene \cite{blees2015graphene}. When strained, this goes through four stages of deformation \cite{Campbell_kirigami_2014, Campbell_kirigami_2016}, of which the first two are primarily in plane while in the latter two, the structure twists out of the plane and starts tearing apart. In the first stage, even while the inter-atomic bond lengths do not change significantly the Kirigami itself expands due to structural cuts. In the second stage, the inter-atomic bonds themselves start expanding and bending which requires the evaluation of the effective fermionic hoppings dynamically \cite{Campbell_kirigami_2016}. On the other hand, cross-cut Kirigamis (also called auxetic structures \cite{tao2023Engineering, rafsanjani2017buckling, lee2020multiaxially, grima2000auxetic, shuaibu2024advancing}) have a larger window of in-plane deformations which are further tunable by controlling the thickness of Kirigami and the size of cuts. In our work, we will work with auxetic Kirigamis and assume the hopping parameters to be static such that all deformations are in the plane. Thus for an undeformed structure ($\theta=0^\circ$) of total length $l = L L_B$, at any angle $\theta$ the total length of the network changes to $l'$ such that \cite{grima2000auxetic}

\beq
l' = l [\cos (\theta) + \sin (\theta)]
\label{eqn:exptheta}
\eeq
where the effective strain can thus be estimated as $\epsilon_s = (l'-l)/l$. It is known from experimental studies on auxetic metamaterials \cite{rafsanjani2017buckling} that depending on the ratio of the thickness and junction width, the system can show in-plane deformations even for significant strains. While the direct translation of these mechanical properties into nanomaterials is far from obvious, approximately $30 \%$ strain, corresponds to $\theta \sim 22^\circ$ (see \eqn{eqn:exptheta}). In graphene systems \cite{blees2015graphene} a strain of $\sim 240 \%$ was realized, however, this necessarily introduces significant inhomogeneities and twisting out of the plane. In most of our study, we will assume homogeneous deformations for analytic control on the physics. Thus it will be applicable when the strain effects are small and the Kirigami structure doesn't encounter any drastic modifications. However, as we will show these results are easily generalizable even in the presence of inhomogeneities and also open up new questions in non-homogeneous systems. 

\section{Multiscale methodology}
\label{sec:method}

We work in a ballistic regime where electronic transport can be estimated via the transmission properties of the network, which can in turn be estimated through a wavepacket evolution. However, as we will illustrate, it is useful both physically and numerically \cite{numcomment} to perform this analysis in three stages: (i) we first characterize the junction} 
{\it between} adjoining blocks such that they are governed by a scattering matrix between the edge states. Any node in this network is composed of two incoming ($1,3$) and two outgoing channels ($2,4$) (see \Fig{fig:method}(a)). The scattering matrix $S$ is defined by two parameters $r$ and $t$ which are in general $\theta$ dependent
\begin{equation}
\begin{pmatrix}
\Psi_2\\
\Psi_4    
\end{pmatrix}
=S
\begin{pmatrix}
\Psi_1\\
\Psi_3    
\end{pmatrix} \textrm{where} \quad S =\begin{pmatrix}
-r&t\\
t&r
\end{pmatrix}
\end{equation}
where $|r|^2+|t|^2=1$ \cite{kramer2005random}. {{An accurate estimation of 
$r,t$ at any node thus completely subsumes its local geometrical structure including any angular dependence, and in general $r,t$ can vary from node to node. (ii) After this first step, we show that the wave-packet dynamics can be modelled via effective non-Hermitian junction (ENHJ) where chiral edge states are replaced by one-dimensional non-Hermitian Hatano-Nelson (HN) models \cite{HatanoPhysRevLett, HatanoPhysRevB} and the nodes are replaced by a four-site scattering junction as shown in \Fig{fig:method}(b). In a one-dimensional Hatano-Nelson (HN) model, fermions are allowed to hop in just one direction which can in turn model the edge states of a quantum Hall state. This uses the mapping of boundary theories of $d$ dimensional topological phases to $d-1$ dimensional non-Hermitian systems  \cite{VishwanathPhysRevLett, SimonLieuPhysRevB, PhysRevLett.120.146402, PRXQuantum.4.030315} since this prevents any backscattering characteristic of a chiral system. Also, any non-homogeneity in the network can be effectively considered within the ENHJ formalism. (iii) In the final step of coarse-graining, one replaces each edge (side) with a single site, forming the median lattice of the network (see \Fig{fig:method}(c)). The non-Hermitian hopping amplitudes on this median lattice effectively model the time-evolution operator akin to the physics of Chalker-Coddington (CC) networks \cite{chalker1988percolation}. This analysis provides an analytic control to understand the phase diagram (see \Fig{fig:squarekiri}(c-d)) in a translationally invariant network. Moreover, this provides a controlled starting point for introducing inhomogeneities in the network. This multiscale approach also leads to numerically inexpensive simulations that can be easily generalized. The total size of simulations progressively changes from $2L^2 L^2_B \rightarrow 4 L L^2_B \rightarrow 4L^2_B$ under each step of coarse graining\cite{numcomment}. In the next section, we discuss the results of these methods as applied to our Kirigami network.}}

\section{Results}
\label{sec:results}

The discussions and results of the three methods as introduced in the earlier section are now discussed in detail. We first characterize the properties of nodes using wave packet dynamics.

\subsection{Wave-packet dynamics }
\label{sec:Wave-packet-dynamics}

We study the junction by first choosing two finite-sized square lattices (describing two Kirigami blocks) each of size $L \times L$ such that the total number of sites is $N=2 L^2$. Every site has two orbitals $A$, $B$ such that the total number of fermionic orbitals is $2N$. The Hamiltonian describing these two blocks separately is $H_{\text{BHZ}}$ and that describing the couplings between these two blocks is $\equiv H_{\text{coupling}}$. Thus, the total Hamiltonian of the system is given by,
\begin{equation}
H=H_{\rm{BHZ}}+H_{\rm{coupling}}
\label{completeH}
\end{equation}

\begin{figure*}
    \centering
    \includegraphics[width=2\columnwidth]{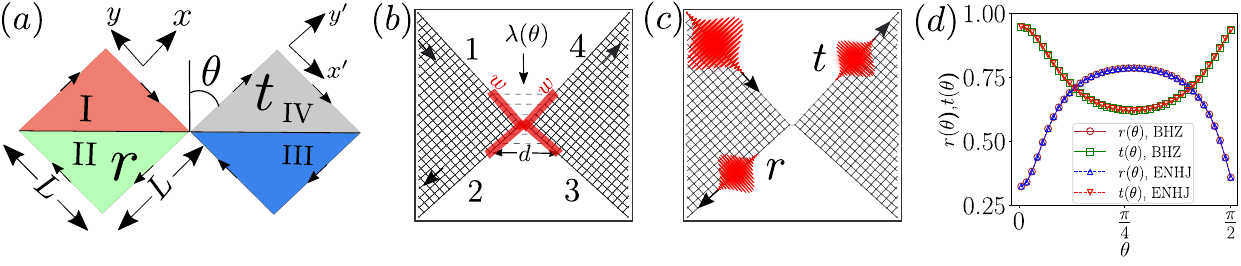}
    \caption{{\bf Model}: (a) A junction is modelled with two square Kirigami blocks meeting at a node. While each block has a size $L\times L$, it has a local coordinate axis that rotates as a function of the deformation angle. The Kirigami blocks can be divided into four regions I-IV to estimate the scattering matrix coefficients (see section~\ref{sec:Wave-packet-dynamics}). (b) Schematic showing that the interblock couplings are limited to a width $w$ and have an angle-dependent hopping of strength $\lambda(\theta)$ between sites separated by a distance $d(\theta)$. (c) Schematic showing that an incident electronic wave packet (from edge channel 1) on a node, gets reflected (on channel 2) and transmitted (on channel 4) by scattering matrix elements $r$ and $t$ respectively. (d) Behavior of $r(\theta), t(\theta)$ as a function of $\theta$ using wave-packet dynamics (see section \ref{sec:Wave-packet-dynamics}) for $M=-3.6,\alpha=1,\beta=1, w=5,N=1250$. Here $N$ is the total number of sites in both the blocks forming the junction. Further, using these obtained $r(\theta),t(\theta)$ for the ENJH model, where the node itself is replaced by four sites, with inter-site non-Hermitian hoppings $r$ and $t$. Further, using the wavepacket dynamics for the ENHJ model, the plot $r(\theta), t(\theta)$ exactly match with those obtained using the BHZ model (see section~\ref{sec:ENHJ}).}
\label{fig:junction}
\end{figure*}

\ul{$H_{\text{BHZ}}$:} Within every Kirigami block the hopping Hamiltonian is given  by
\beq
H_{\text{BHZ}} = \sum_{ i, \delta } \Psi^\dagger_i T_{\delta} \Psi_{i+\delta} + \text{h.c.} + \sum_i \Psi^\dagger_i \Gamma \Psi_{i}
\eeq
where $\delta=\hat{x}, \hat{y}$, the unit vectors in the $x$ and $y$ directions defined with respect to a local axis defined for every Kirigami block (see \Fig{fig:junction}(a)). $\Psi_i \equiv \Big(c_{iA}, 
c_{iB} \Big)^T$ where $c_{iA}, 
c_{iB}$ are the fermionic annihilation operators corresponding to the two orbitals $A, B$ at site $i$. $\Gamma$ describes the onsite energies and $T_{\hat{x}}$ and  $T_{\hat{y}}$ are the hopping matrices given by: 
\beq
T_{\hat{x}} = -\frac{1}{2}(\sigma_z+i\sigma_x), T_{\hat{y}} =-\frac{1}{2}(\sigma_z+i\sigma_y), \Gamma =(M+2)\sigma_z
\eeq
where $\sigma_x, \sigma_y, \sigma_z$ are the Pauli matrices.  In a translationally invariant system, this Hamiltonian in the momentum space gives rise to \eqn{BHZham}. Note that as the angle $\theta$ is changed, such that the local coordinate system of every Kirigami block obtains a relative angle with respect to another Kirigami block, the hopping amplitudes {\it within} any Kirigami block does not change. For instance as shown in \Fig{fig:junction}(a) the relative angle between the two Kirigami blocks with local coordinates $(x,y)$ and $(x',y')$ respectively has a mutual angle of $2\theta$.

\ul{$H_{\text{coupling}}$:  } Next we couple the two Kirigami blocks with interblock hoppings as follows. Labeling the sides (edges) forming the junction ($s_e$=1-4) as shown in \Fig{fig:junction}(b), we identify the sites along the edges $n=1,2, \ldots w$. Thus $w$ number of sites, denoting the width of the junction, are coupled between the sides facing each other (for instance, in \Fig{fig:junction}(b), $s_e=1$ and $s_e=4$ and similarly $s_e=2$ and $s_e=3$). Labeling the fermionic creation operators on the edge $s_e$ at site $n$ as $\Psi^\dagger_{n, s_e}$, the $H_{\rm{coupling}}$ is therefore given by
\beq
\hat{H}_{\rm{coupling}}=\sum_{n=1}^{w}  \lambda(n, \theta)  \Big( \Psi^\dagger_{n,1} T_{\hat{x}} \Psi_{n,4} + \Psi^\dagger_{n,2} T_{\hat{x}} \Psi_{n,3} \Big) + \text{h.c.}
\eeq
The functional form of $\lambda(n,\theta)$ implements the interblock hopping amplitudes which we describe next. Given the two sites $n$ on $s_e=1$ and $s_e=4$ the distance between them is $d=2n \sin(\theta)$, and similarly for those between $s_e=2$ and $s_e=3$ is $d=2n \sin(\frac{\pi}{2}-\theta)$. Given a distance $d$ between the two facing sites, the hopping amplitude is given by
\beq
\lambda(\theta) = \alpha \exp(- \beta d(\theta)) 
\eeq
where $\alpha$ and $\beta$ are tuning parameters. We set  $\alpha=1, \beta=1$ unless otherwise stated. The choice of parameters and the coupling Hamiltonian has been kept such that at $\theta=0$, $\lambda(\theta) =1$ thus the coupling reduces to the bare hopping scale of the rest of the BHZ system. 

Having described the Hamiltonian of the system, we next discuss the wavepacket evolution and estimation of $r(\theta)$ and $t(\theta)$. Given a choice of $M$ and $\theta$, we first diagonalize $H$ (see \eqn{completeH}) to obtain the complete eigenspectrum. In the gapped topological regime, the bulk gap in the periodic boundary condition is given $\Delta_g(M)$. We form a projector out of the single particle wavefunctions $|m\rangle$ such that their eigen energies $\epsilon_m$ are $ |\epsilon_m| < \Delta_g(M)$.  The projector is thus given by
\beq
\hat{P} = \sum_{m, |\epsilon_m|<\Delta_g(M)} |m\rangle \langle m |
\label{projec}
\eeq
This projects any wavefunction to the edge state manifold of the complete system.

\begin{figure*}
\includegraphics[width=2.0\columnwidth]{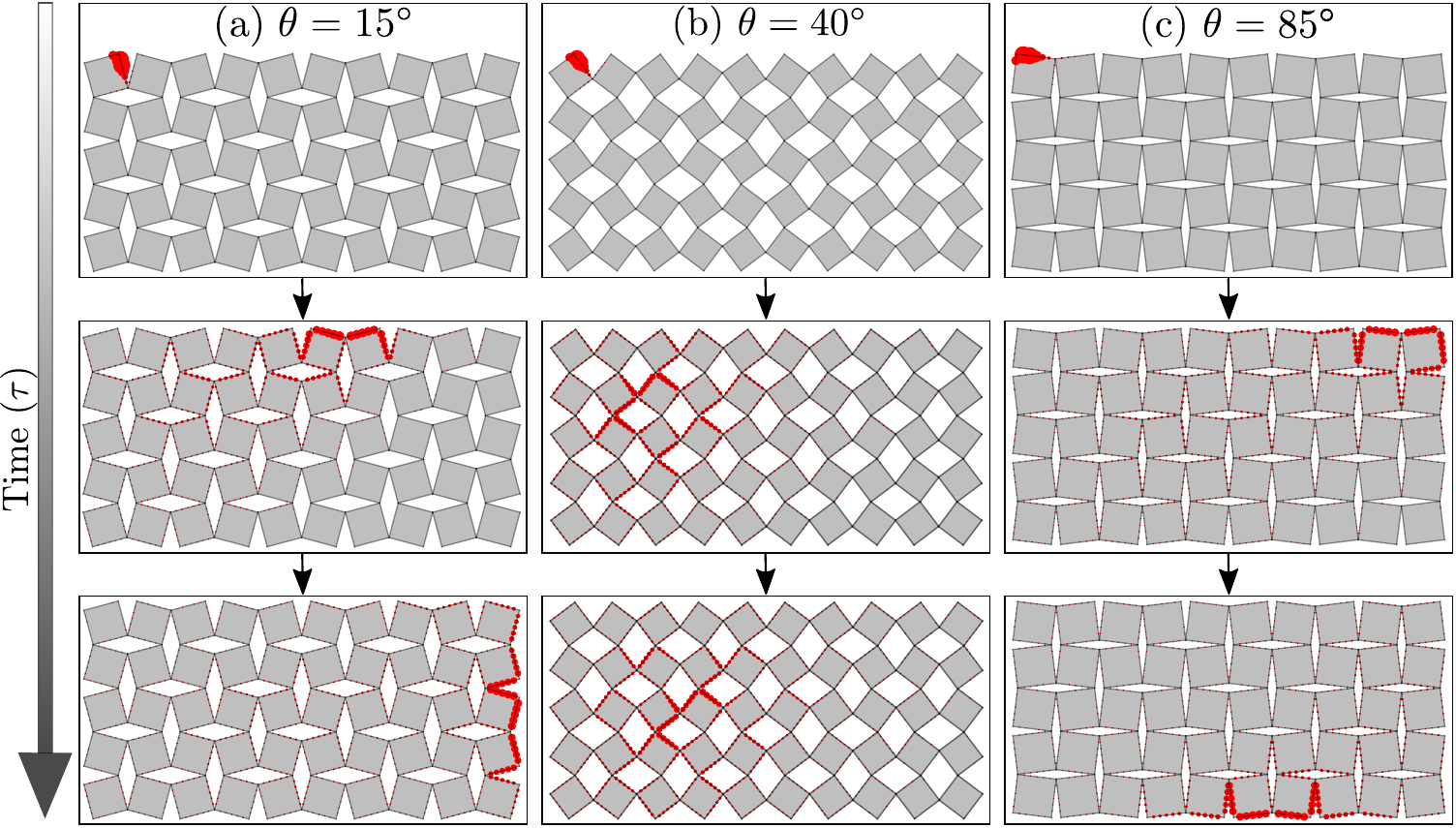}
\caption{{\bf Topological mechanical insulator:~} The time evolution of a wavepacket under the ENHJ formalism (see section.~\ref{sec:ENHJ}) for three configurations of the Kirigami networks at (a) $ \theta=15^{\circ}$ for three different times $\tau=\{2,60,120\}$ (shown by arrows). (b) $\theta=40^{\circ}$ and $\tau=\{2,70,90\}$ (c) $\theta=85^{\circ}$ and $\tau=\{2,90,180\}$. While (a) and (c) show that the network is topological (edge states), in (b) one finds that the network is trivial.}
\label{fig:kiri_wp}
\end{figure*}

To start the wavepacket evolution we initialize a unit state on the top corner of side 1 on the A orbital. We project this state using the projector above (see \eqn{projec}) and renormalize it to obtain the initial state $|\psi_{\text{in}} \rangle$. This state is further evolved using the Hamiltonian (see \eqn{completeH}) to a later time such that final state $|\psi_{\text{fi}} \rangle $ at time $\tau$ is given by
\beq
|\psi_{\text{fi}}\rangle  = \exp(-i H \tau )  |\psi_{\text{in}} \rangle 
\eeq
The initial packet moves towards the junction and scatters into edges 2 and 4. Given the unidirectional chiral edge states, there is neither any back-scattering nor any amplitude on edge 3. After a time when the packet has scattered from the junction, we calculate the total probability density on sides 2 and 4. To estimate the reflected probability density we evaluate 
\beq
p_2= \sum_{i \in \rm{region~II}, \gamma={A,B}} |\langle i \gamma |\psi_{\text{fi}} \rangle |^2
\eeq
and similarly the transmitted probability density is given by 
\beq
p_4= \sum_{i \in \rm{region~IV}, \gamma={A,B}} |\langle i \gamma |\psi_{\text{fi}} \rangle |^2
\eeq
The various regions (I-IV) are shown in \Fig{fig:junction}(a). Consistent with our expectations we find that the corresponding quantities in regions III and I are zero ($p_3=p_1=0$) after the scattering event. Having estimate $p_2$ and $p_4$, we calculate  $r (\theta) = \sqrt{p_2}$ and $t (\theta) = \sqrt{p_4}$.

Choosing $M=-3.6$ in the topological phase with Chern number $C=-1$ we evaluate $r,t$ for all values of $\theta$. We find an intriguing dependence, where $t(\theta) \sim 1$ near both $\theta \sim 0$ and $\sim \frac{\pi}{2}$  while $r(\theta) > t(\theta)$ near $\theta \sim \frac{\pi}{4}$  (\Fig{fig:junction}(d)). We have checked that this behavior is independent of the system size (see Appendix~\ref{sec:appenpara}). {{The choice of parameter values, in particular $w \sim 5$ is kept to ensure a competition with the localization length $\xi$ which changes as a function of $M$. A large value of $w \gg \xi$ will ensure $t\sim 1$ at all values of $\theta$. }} We also note that the exact values depend on the choice of $\lambda(\theta), w$, and $M$, however, the qualitative behavior of these coefficients do not (see Appendix~\ref{sec:appenpara}). With this the first step of coarse-graining (see \Fig{fig:method}) is complete and we move to the ENHJ formalism to investigate the properties of the full network.

\subsection{Effective non-Hermitian junction}
\label{sec:ENHJ}
{{In order to study the transport property of the full network - we devise an ENHJ formalism where the scattering properties of every node are replaced by a non-Hermitian scatterer as seen in \Fig{fig:method}(b). This effectively ensures no back-scattering and with a given value of $r,t$ exactly reproduces the reflection and transmission probabilities through the junction (see \Fig{fig:junction}(d)). For the complete network, a non-Hermitian Hamiltonian is thus given by
\beq
H = H_{\text{edge}} + H_{\text{hyb}}
\label{eq:ENHJ}
\eeq
where
\beq
H_{\text{edge}} =  \sum_{i,s_e} -  c^\dagger_{\vec{r}_i+ \hat{e}, s_e} c_{\vec{r}_i,s_e} 
\eeq
specifies the edge Hamiltonians. $s_e$ specifies any edge with the chiral current in direction $\hat{e}$ and $\vec{r}_i$ are the discrete positions labeling the sites on any edge.
$H_{\text{hyb}}$ represents the mixing between edge modes at two types of nodes $n \in \{ N_1, N_2\} $ (having opposite arrow directions, as shown in \Fig{fig:method}(b), with yellow and blue colors respectively), such that 
\bea
&H_{\text{hyb}}& = \sum_{n \in N_1} \Big( [-r d^\dagger_{2,n} + t d^\dagger_{4,n}   ] d_{1, n} +  [ t d^\dagger_{2,n} + r d^\dagger_{4,n} ]  d_{3, n} \Big) \notag \\
&+& \!\! \sum_{n \in N_2} \Big( [r d^\dagger_{3,n} - t d^\dagger_{1,n}   ] d_{2, n} +  [ t d^\dagger_{3,n} + r d^\dagger_{1,n} ]  d_{4, n} \Big) 
\eea
where $s_e=$1-4 are the edges labeled anticlockwise at any node, and edges $\{1,3\}$ and  $\{2,4\}$ have the incoming (outgoing) and outgoing (incoming) edge currents in nodes $N_1 (N_2)$.  $d^\dagger_{s_e,n}$ represents the fermionic operators (identical to $c^\dagger$ operator) of the edge $s_e$ which forms part of the four-site junction, with $r,t$ being the scattering amplitudes of any node $n$.  Given the complete morphology of the Kirigami network, every node, in general, can have different values of $r,t$ decided by local junction parameters such as $\theta, w$~etc. 

}}

{{
With this coarse-graining step, we can now pose how the complete network behaves given a value of $r,t$ for each node. To make progress we first study the case where $r_n, t_n = (r,t)$ independent of node position, thus leading to a translational symmetry on the network. We first study the Kirigami in the relatively little strain ($\theta=  15^\circ$) which corresponds to $r,t =(0.52, 0.85)$. We again release a wavepacket on the edge of the network and evolve using the ENHJ Hamiltonian (see~eqn.\ref{eq:ENHJ}), we find that the wavepacket evolves and moves through the boundary of the network reflecting the topological character of the full network consistent with topologically non-trivial Kirigami blocks (see \Fig{fig:kiri_wp}(a)). When the Kirigami is now stretched to an angle $\theta \sim 40^\circ$ with $r,t =(0.78, 0.63)$, we notice something surprising. A similar wave packet now scatters into the bulk showing no protection to the edge state, thus suggesting that the complete network is trivial even though each of the blocks is topological (see \Fig{fig:kiri_wp}(b)). A further increase in $\theta = 85^\circ$ again makes the network topological (see \Fig{fig:kiri_wp}(c)). This reflects the central result of this work, where a mechanical deformation leads to a drastic change in the transport properties of the system. We have further checked that realistic gradients in strain even up to $\sim 15-30\%$ or so do not change these properties (see Appendix.~\ref{sec:stretch}). The above analysis also suggests the existence of a critical angle, at which the network must transit from a topological to a trivial phase. To uncover such a transition we do further coarse-graining and analyze the system within a Chalker-Coddington formalism.}}

\subsection{Chalker Coddington networks}
\label{sec:CC}
Having defined the property of every junction and the observation of strain dependent transport of the full network we now pose if there is a critical angle where the transition from topological to trivial phase occurs for the complete Kirigami network. A Chalker-Coddington framework is particularly useful to analyze this\cite{chalker1988percolation}. Briefly, the network can be studied within a transfer matrix formalism where an initial state evolves unitarily as \cite{GruzbergPRBkagome} 
\beq
|\Psi_f\rangle =  U_{\text{eff}}  |\Psi_i\rangle
\eeq
where the effective evolution operator, $U_{\text{eff}}$ is decided by the scattering matrix structure of the network. The $U_{\text{eff}}$ is constructed using a median lattice where each edge of the Kirigami block is replaced by a site such that the system has a 16-site unit cell. The `hopping' elements are represented by $r(\theta)$ and $t(\theta)$ with appropriate convention depending on the direction of the edge current. For analytic control we assume that the network is translationally invariant and see if it retains any edge states we set it up on a ribbon geometry which is finite (periodic) in the $y$ ($x$) direction. We find the eigenvalues of 
$U_{\text{eff}}$ which are of the form $e^{-i \epsilon}$ and plot $\epsilon$ as a function of $k_x$. Note that $r,t$ enter as parameters in the transfer matrix. The network description closely follows the formalism developed in Charles~et.~al.\cite{GruzbergPRBkagome}, done in the context of Chalker-Coddington networks for integer quantum Hall transition.

\begin{figure}
\includegraphics[width=0.9\columnwidth]{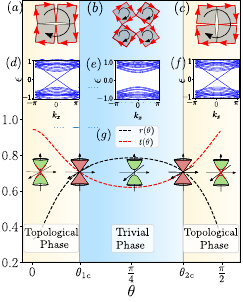}
\caption{{\bf Network model:~}(a)-(c) Schematic of Kirigami network as function of angle for $\theta \sim 0, \frac{\pi}{4}$ and $\frac{\pi}{2}$. While each block contains edge states (red arrows), they hybridize to render the complete network topological in (a) and (c), and trivial in (b). (d)-(f) Behavior of eigenvalues $\epsilon$ of $i \ln (U_{\text{eff}})$ as a function of $k_x$ then the network is solved on a ribbon geometry where $y$ direction is kept open. Edge states are present when $0 < \theta < \theta_{1c}$ (d) and $(\theta_{2c} < \theta < \frac{\pi}{2})$ (f) while the system is trivial between $(\theta_{1c} < \theta < \theta_{2c} )$ (e). (g) The behavior of $r(\theta)$ (black dashed line) and $t(\theta)$ (red dashed line) as a function of $\theta$ for $N=1250,M=-3.6, \alpha=1,\beta=1,w=5$ along with schematic figures showing the bulk gap closings. Dirac cone closings happen  at $\theta_{1c}$ and $\theta_{2c}$ when $r(\theta)=t(\theta)$.}
\label{fig:fig3}
\end{figure}

\begin{figure}
\centering
\includegraphics[width=0.8\columnwidth]{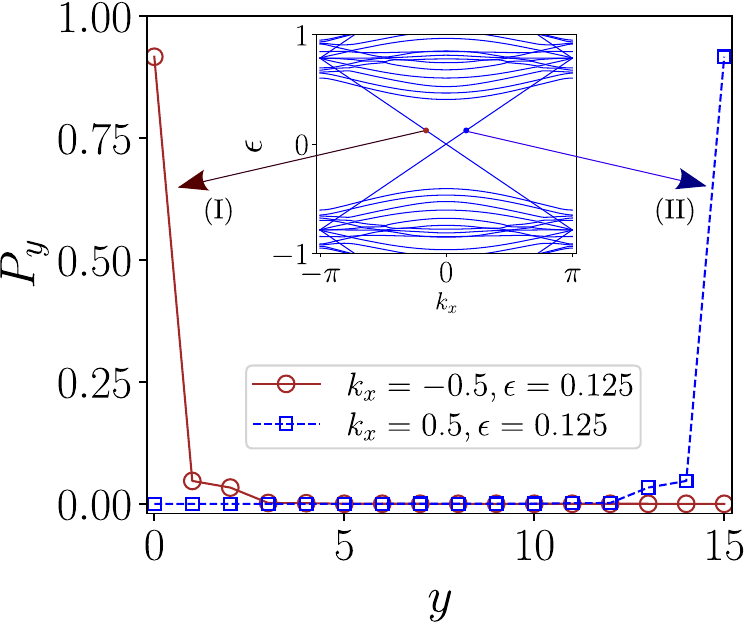}
\caption{{\bf Edge state profile:~} $P_y$ as function of $y$ when $L_y=16, r=0.4, t=0.916$. For (I) $k_x=-0.5, \epsilon=0.125$ and for (II) $k_x=0.5, \epsilon=0.125$ }
\label{fig:Edge_state}
\end{figure}

When $r<t$ and $\theta < \theta_{1c}$, we find linearly dispersing edge states on the boundary (see \Fig{fig:fig3}(d)) reflecting that the topological character of the individual Kirigami block also leads to an overall chiral edge in the system (\Fig{fig:fig3}(a)). 
{{The electronic probability density $P_y$ (as a function of $y$) for given values of $k_x$ and $\epsilon$ is shown in \Fig{fig:Edge_state}. As can be seen, the right (left) moving edge state resides on the bottom (top) of the ribbon. These states are absent when periodic boundary conditions are used in the $y$ direction.}} However now as $\theta$ is increased, or the structure opens up, the individual scattering elements ($r(\theta), t(\theta)$) modulate approaching a critical point $\theta_{1c}$ where $r=t$. At this point, the system undergoes a phase transition where the
$\epsilon$ shows a Dirac-like gap closing. After $\theta>\theta_{1c}$
the gap opens up again
(\Fig{fig:fig3}(e)) with no chiral mode on the complete network. This reflects that the Kirigami network has become a ``trivial" insulator (see \Fig{fig:fig3}(b)). This is an example of a mechanical deformation-induced topological phase transition. {{In order to corroborate the results from ENHJ formalism and CC networks we compare the spectrum and show in Appendix.~\ref{sec:corres}.}}
With further increase in $\theta$ another critical point $\theta_{2c}$ is approached where the system transits back from the trivial to topological phase (see \Fig{fig:fig3}(c) and (f)). The specific values of $\theta_{1c}$ and $\theta_{2c}$ are dependent on both the microscopic parameters of the Hamiltonian governing each block, as well as the nature of the network. For instance in the case of the square network shown in \Fig{fig:squarekiri}(a), $\theta_{1c}$ and  $\theta_{2c}$ are symmetric about $\pi/4$. For every value of $M$ numerically evaluated $r=t$ points leads to the phase diagram as shown in \Fig{fig:squarekiri}(c),(d). While we have chosen $r,t$ to be real, a uniform phase does not change $\theta_{1c},\theta_{2c}$. However relative scattering phases either uniform or randomly varying from block to block can lead to both changes in critical angles and new phases \cite{GruzbergPRBkagome} which we do not explore here.

We note that this analysis and formalism is only applicable when each of the Kirigami blocks resides in the topological regime i.e. $-4<M<0$. For other values of $M$ when each of the blocks is itself an insulator - mechanical deformation doesn't render it topological. Each block has a topological phase transition within $-4<M<0$ at $M=0,-2$, and $-4$ where the system contains Dirac cones. Here again, the blocks are themselves metallic and this analysis breaks down. Given the $M=-2$ point sits between the two topological Chern insulator phases of Chern number $\pm 1$, $r=t$ by symmetry reasons. Overall the phase diagram is further symmetric about $M=-2$ since apart from the direction of the chiral edge states the physics remains the same.

It is interesting to point out the physics of the $r=t$ line in this system, particularly near $\theta \sim 0$. Given when each Kirigami block is in the topological regime - there exists an edge state with an edge localization length $\xi$, which is dependent on the bulk gap which in turn depends on $M$. Thus whether an edge wavepacket can transmit or reflect from the junction depends on whether $\xi \ll w$ or $\xi \gg w$ respectively. Therefore given a choice of $w$, there exists a critical value of $M$ even within the topological phase, defined as $M^*$ at which $\xi(M) \sim w$, therefore leading to a point where $r=t$. This value of $M^*$ is thus independent of the system size of the Kirigami block but only dependent on the value of $\theta$ and the parameters defining the junction (see Appendix~\ref{sec:appenMstar}). We further note tuning $\alpha, \beta$ which defines the strength of the junction can further tune the values of $r,t$ (see Appendix~\ref{sec:appenpara}).

\section{Triangular Network}
\label{sec:triang}

\begin{figure}
\includegraphics[width=0.9\columnwidth]{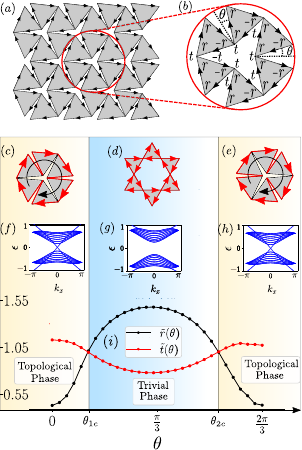}
\caption{{\bf Triangular network Kirigami:~} (a) Schematic of the Kirigami network made up of triangular blocks and the corresponding convention of scattering matrix elements is shown in (b). (c)-(e) Schematic of the network at $\theta \sim 0, \frac{\pi}{3}$ and $\frac{2\pi}{3}$. Each triangular block containing the edge states (red arrows) corresponds to the topological phase of the network as shown in (c) and (e), and the trivial phase in (d). The ribbon geometric calculation of the network with periodic in $x$ and open in $y$ direction shows the energy eigenvalues as a function of $k_x$ in (f)-(h). The edge states are present when $0 < \theta < \theta_{1c}$ (f) and $(\theta_{2c} < \theta < \frac{2\pi}{3})$ (h) while the system is trivial between $(\theta_{1c} < \theta < \theta_{2c} )$ (g). (i) The behavior of $\tilde{r}(\theta) $ (black line) and $\tilde{t}(\theta)$ (red line) as function of $\theta$ for $N=256,M=-3.5,\alpha=2,\beta=1,w=3$.}
\label{fig:fig4}
\end{figure}

In order to test these physics in another network, we set up the same problem however now on a Kirigami network where the network deforms between the triangular lattice to a Kagome-like network as a function of deformation angle $\theta$ (see \Fig{fig:fig4}(a),(b)). The microscopic Kirigami block is still assumed to be governed by a square lattice BHZ model. Given a different network structure the $U_{\text{eff}}$ changes. However again as a function of $\theta$ the system undergoes topological to trivial to topological phase transitions as shown in (see \Fig{fig:fig4}(c)-(e)). What is interesting however is that the critical points $\theta_{1c}$ and $\theta_{2c}$ are now symmetric about $\theta=\frac{\pi}{3}$. Moreover the topological transition occurs at $r(\theta) = \frac{1}{2}, t(\theta)= \frac{\sqrt{3}}{2}$. Thus the transitions can be found when $\tilde{r}(\theta) = \tilde{t}(\theta)$ where  $\tilde{r}(\theta) \equiv 2 r(\theta)$ and  $\tilde{t}(\theta) \equiv \frac{2}{\sqrt{3}} t(\theta)$. The behavior of $\tilde{r}(\theta)$ and $\tilde{t}(\theta)$ for $M=-3.5$ is shown in \Fig{fig:fig4}(i). The spectrum when evaluated on a ribbon geometry clearly shows the edge states in the topological regimes (see \Fig{fig:fig4}(f-h)). 
This reflects the generality of this physics to various Kirigami systems and that critical angles for transitions can themselves be tuned by nature of the Kirigami structure. 

\section{Discussion and Summary}
\label{sec:discussion}

Metamaterials and mechanically engineered systems herald the new era of technological devices. A host of electronic systems based on such structures have been realized. However, the interplay of quantum phases and their transitions vis-a-vis their mechanical tunability has been little explored. In this work, we show that topological phases, in particular a Chern insulator which contains chiral edge channels can be tuned to a trivial phase by such mechanical tuning. Recent experimental progress, in particular, such as Kirigami-based systems \cite{zhai2021mechanical, chen2020kirigami, LishuaiADvMat2023} have significantly expanded the capabilities to realize this physics.

{{We comment on the parameter values where our results are applicable. The quantum coherence effects of mechanical deformations etc.~can be experimentally seen only when system length scales are comparable to electronic decoherence length scales. Various realistic simulations with and without molecular dynamics (MD) in graphene/hBN samples/diamond thin-films where effects of mechanical deformation on electronic structure have been studied work with Kirigami blocks where typical $L \sim 50$\AA, $w \sim 20$\AA ~and $L_B \sim 10$ which is at the same parameter regimes where we have performed our analysis \cite{Campbell_kirigami_2016, dey2024enhancing, han2017super, kumar2020manipulation,zhu2024mechanical}. We believe therefore realistic MD and density functional theory (DFT)-based simulations can be performed on layered topological insulating materials such as Bismuthene \cite{Pezo_PRM_2021, reis2017bismuthene, focassio2021structural}, HgTe \cite{Kozlov_PRL_2014, Olsh_PRL_2015, kvon2020topological} to find direct signatures of such deformation-dependent quantum transport. Experiments in graphene Kirigamis \cite{blees2015graphene} and metamaterials \cite{el2022mechanical, LishuaiADvMat2023, tao2023Engineering, liu2022robust, tang2015design} yet, are considerably larger in sizes such that one expects transport to have significant decoherence effects. These results will also be applicable to twisted bilayer graphene systems where the transport is dominated by edge currents on effective triangular networks \cite{Kundu_PRB_2024, yoo2019atomic, Huang_PRL_2018, BeuleRPL2020, VakhtelPRB2022}.  Our results therefore motivate further experimental work in the direction of Kirigami-inspired quantum materials.}}

To summarize, our study places the idea of Kirigami inspired topological phases and their transitions on a concrete footing, uncovering their mechanism in a simple setting. Further studies on effects of realistic strains, gradients in deformations, strong correlations, and interactions are some of the many future directions this work opens up. Equivalent physics in alternate platforms such as photonics and phononic systems may be another interesting direction to pursue.

\section{Acknowledgement} 
We acknowledge fruitful discussions with G. Murthy, G. Sreejith, Diptarka Das, Arijit Kundu, Subrata Pachhal, Sudipta Dubey, Ritajit Kundu, Saikat Mondal and Soumya Sur. RS acknowledges funding from IIT Kanpur Institute Fellowship. AA acknowledges support from IITK Initiation Grant (IITK/PHY/2022010).

\section{Appendix}
\label{sec:appen}

\appendix

\begin{figure}
\centering
\includegraphics[width=0.6\columnwidth]{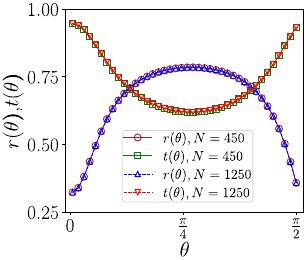}
\caption{{\bf System size ($N$) dependence:~} The plot of  $r(\theta)$ and $ t(\theta)$ in BHZ junction modelling (see Fig.\ref{fig:junction}) when $N=450$ and $N=1250$ with $\alpha=1, \beta=1, w=5$ and $M=-3.6$.}
\label{fig:system_size}
\end{figure}

\begin{figure}
\centering
\includegraphics[width=1.0\columnwidth]{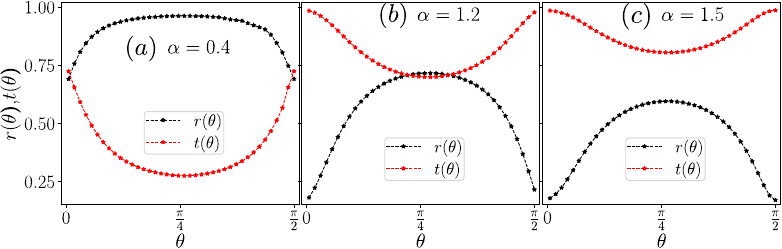}
\caption{{\bf Coupling dependence:~} $r(\theta)$ and $ t(\theta)$ in BHZ junction modelling (see Fig.\ref{fig:junction}) with ($N=1250$, $M=-3.6, \beta=1$ and $w=5$). (a) $\alpha=0.4$ corresponds to the $r=t$ when $\theta \approx 0$ or $\frac{\pi}{2}$ and $r\approx 1$, $t\approx 0$ when $\theta \approx \frac{\pi}{2}$. (b) For $\alpha=1.2$, $r=t$ when $\theta \approx \frac{\pi}{4}$.  (c) For $\alpha=1.5$, $t(\theta) > r(\theta)$. }
\label{fig:parameters}
\end{figure}
\begin{figure}
\centering
\includegraphics[width=1.0\columnwidth]{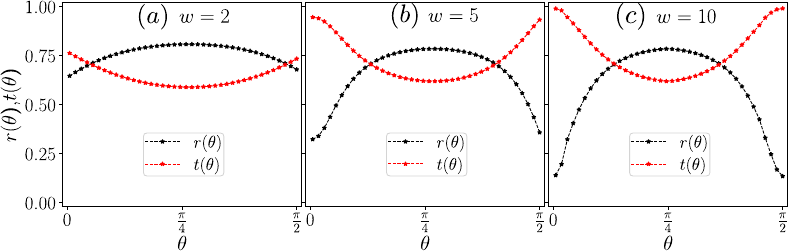}
\caption{{\bf $w$ dependence:~}The junction width $(w)$ dependence of $r(\theta)$ and $ t(\theta)$ in BHZ junction modelling (see Fig.\ref{fig:junction}) with ($N=1250$, $\alpha=1, \beta=1$ and $M=-3.6$). }
\label{fig:w_dependence}
\end{figure}

\begin{figure}
\centering
\includegraphics[width=1.0\columnwidth]{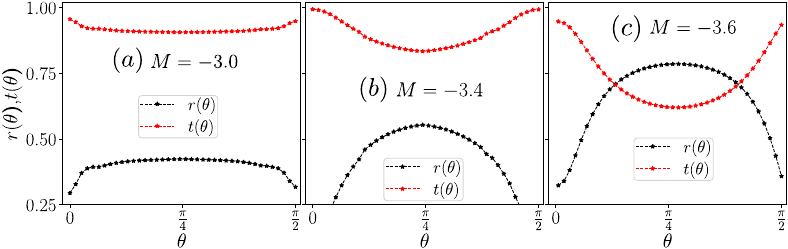}
\caption{{\bf $M$ dependence:~}$M$ dependence of $r(\theta)$ and $ t(\theta)$ in BHZ junction modelling (see Fig.\ref{fig:junction}) with ($N=1250$, $\alpha=1, \beta=1$ and $w=5$).}
\label{fig:M_dependence}
\end{figure}

\section{Parameter dependence of $r(\theta)$ and $t(\theta)$}
\label{sec:appenpara}

In this section, we investigate and present additional results on the functional dependence of $r(\theta)$ and $t(\theta)$ on various parameters such as its system size ($N$), BHZ mass parameter $M$, width $w$ and coupling parameter $\alpha$.

\ul{System size ($N$) dependence: } 
With the fixed set of parameters $(\alpha,\beta,w$ and $M$), we model the junction using both the microscopic BHZ junction modelling (\eqn{completeH}) and the ENHJ modelling (\eqn{eq:ENHJ}). With the wavepacket evolution, we find that the functional forms of $r(\theta)$ and $t(\theta)$ are independent of the system size ($N$) as shown in \Fig{fig:system_size} (not shown for ENJH model).

\ul{Coupling dependence $(\alpha, w)$: } Given a set of parameters ($N, M, w$, and $\beta$) we find that tuning the coupling strength ($\alpha$) can lead a quantitative change  of  $r(\theta)$ and $t(\theta)$, as illustrated in \Fig{fig:parameters}. The smaller value of $\alpha$ reduces the junction strength impeding transmission thus resulting in $r(\theta)>t(\theta)$. Interestingly, we note that for $ \alpha=0.4$, the junction shows $r=t$ when $\theta \approx 0$ and $\frac{\pi}{2}$ (\Fig{fig:parameters}(a)). This corresponds to the critical angles $(\theta_{1c}$ and $\theta_{2c})$. By tuning $\alpha$ to $1.2$ one can tune these critical angles up to  $\theta \approx \frac{\pi}{4}$ (\Fig{fig:parameters}(b). Further, increase in $\alpha$ makes $t(\theta)>r(\theta)$ as shown in \Fig{fig:parameters}(c). Similarly tuning $w$ can allow further modulation of $r(\theta)$ and $t(\theta)$ as shown in \Fig{fig:w_dependence}. Thus $\alpha, w$ serve as important tuning parameters in the system.

\ul{$M$ dependence: } The localization length {$\xi$} of the edge states is related to $M$. Deep in a topological phase when $M=-3$ and a given $w=5$ the system has $\xi \ll w$ thus it allows easy transmission with $(t>r)$ as shown in \Fig{fig:M_dependence}(a) for all values of $\theta$. Note as $M$ is decreased further to $M=-3.6$, $\xi(M)$ increases since the bulk gap closing happens at $M=-4$. Given $t \sim 1$ near $\theta \sim 0$ implies that $\xi(M) < w$. However now in an intermediate range of $\theta$
near $\frac{\pi}{4}$ we find $r>t$. This increase in $r$ can be attributed to an effective reduction of junction coupling due to a coupling strength ($\lambda(\theta)$). 

{{

\section{Stretched Kirigami and non-uniform strain}
\label{sec:stretch}

\begin{figure*}
\centering
\includegraphics[width=2\columnwidth]{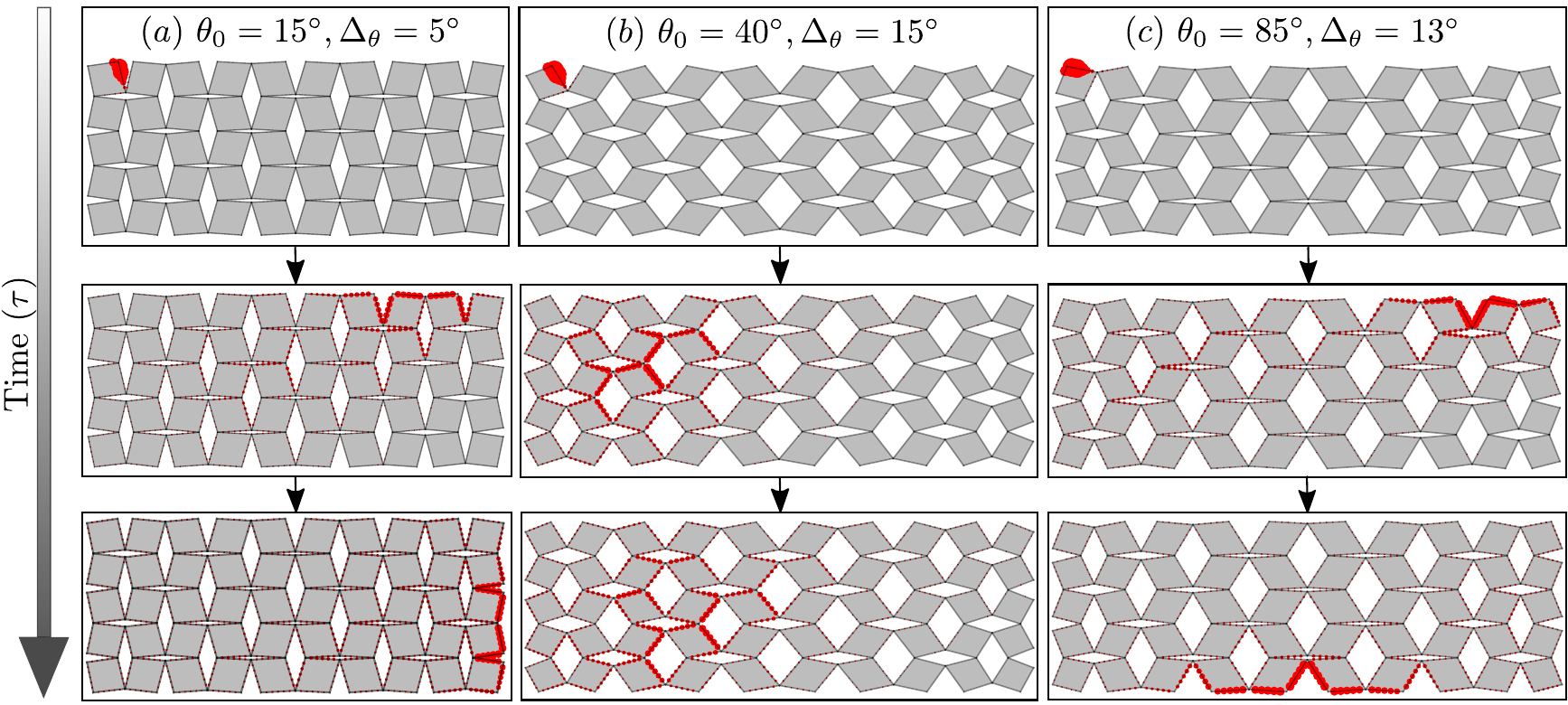}
\caption{{\bf Stretched Kirigami:~} The time evolution of wavepacket for three configurations of the Kirigami networks at (a) $\theta_o =15^{\circ}, \Delta_\theta= 5^{\circ} $ for three different times $\tau=\{2,80,130\}$ (shown by arrows). (b) $\theta_o = 40^{\circ}, \Delta_\theta= 15^{\circ}$ and $\tau=\{2,60,80\}$. (c) $\theta_o = 85^{\circ}, \Delta_\theta=  13^{\circ}$ and $\tau=\{2,80,190\}$.}
\label{fig:stretchKiri}
\end{figure*}

In order to study the effects on realistic strain, as seen in experiments and simulations \cite{an2020programmable}, we model a strain gradient and see its effect on the transport properties. We assume translational invariance in the vertical ($y$) direction and assume that a node $n$ at any position $x_n \in \{-\frac{l'}{2},\frac{l'}{2}\}$ where $l'$ is the size of the complete Kirigami network at an angle $\theta_o$ (see \eqn{eqn:exptheta}). To include strain we further add a Gaussian profile such that any node has an angle $\theta_n = \theta_o - \delta \theta_n$ where

\beq
\delta \theta_n  = \frac{\Delta_\theta}{(e^{-1/2}-1)} \left[\exp \left( - \frac{2x_n^2}{l'^2} \right) - 1 \right]
\eeq

This ensures that as we move from the center to the edge of the Kirigami networks the node angles smoothly deform from $\theta_o$ to $\theta_o - \Delta_\theta$ as seen in experiments on auxetic Kirigamis \cite{}. Given the characterization of the node (see \Fig{fig:junction}) a local angle of $\theta_n$ corresponds to a scattering coefficient $\{r_n,t_n\}$. Thus $\{r_n, t_n\}$ will vary from $\{r(\theta_o), t(\theta_o)\}$ at the center to $\{r(\theta_o - \Delta_\theta), t(\theta_o - \Delta_\theta)\}$ at the edges of the Kirigami network. 
We denote the maximum deviation in $r_n$ from center to edge as $\Delta_r=  r(\theta_o) -  r(\theta_o - \Delta_\theta). $ Assuming $\Delta_\theta \ll \theta_o$ we linearly extrapolate the variation of $r_n$ to be 
$r_n= r(\theta_o) - \delta r_n$ where

\beq
\delta r_n  = \frac{\Delta_r}{(e^{-1/2}-1)} \left[\exp \left( - \frac{2x_n^2}{l'^2} \right) - 1 \right]
\eeq

such that at every node $|t_n| = \sqrt{1- r^2_n}$. 

In \Fig{fig:kiri_wp} we had shown the character of the network when the node angles were uniform i.e.~$\theta_o = 15^{\circ}, 40^{\circ}, 85^{\circ} $. We now add a $\Delta_\theta = 5^{\circ}, 15^{\circ}, 13^{\circ} $ in each of the cases and show the wavepacket evolution in \Fig{fig:stretchKiri}.  This shows that the results of the uniform strain case is stable to small gradients in the strain profile. This is however only true deep in the topological/trivial phase such that 
local $r_n$ at any node do not cross the phase boundaries. Strongly non-uniform strains ($\Delta_\theta \gg \theta_o$) may bring interesting transport properties which is an interesting future work.

\section{Correspondence between ENHJ and CC networks}
\label{sec:corres}

In order to further corroborate the results between the ENHJ and the CC networks we compare the spectrum of the ENHJ and the CC network in a strip geometry with different values of $r$ in \Fig{fig:energy_vs_r_in_real_space}. For a given choice of $r(\theta), t(\theta)$ we find the eigenvalues of the evolution operator constructed using ENHJ formulation ($U_{\text{NH}}$) which are the form $e^{-i\chi}$ and plot $\chi$ as a function of $r$. With open boundary conditions, we find that close to $\chi = 0 $ and $r<r_c=\frac{1}{\sqrt{2}}$ the plot has ``edge states" as shown in Fig.~\ref{fig:energy_vs_r_in_real_space}(a). These edge states disappear under the periodic boundary conditions as shown in Fig.~\ref{fig:energy_vs_r_in_real_space}(b). This clearly shows the presence of edge states and the topological character in the network. When $r>r_c= \frac{1}{\sqrt{2}}$, there are no edge states also, $\chi$ has a finite gap, which shows the trivial insulating phase of the network. Clearly, ENHJ formulation is also able to capture the topological phase transition and presence of critical point $r_c= \frac{1}{\sqrt{2}}$  in the CC network as discussed in section \ref{sec:CC}.

\begin{figure}
    \centering
\includegraphics[width=1\columnwidth]{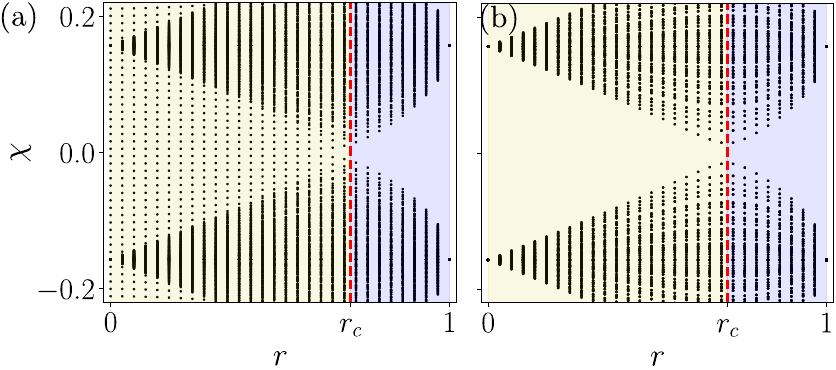}
\caption{{\bf Variation of $\chi$ with $r$:~} Plot of eigenvalues $\chi$ of $i\ln(U_{\text{NH}})$ as a function of $r$ with (a) open boundary conditions, there are edge states for $r<r_c=\frac{1}{\sqrt{2}}$. (b) With the periodic boundary conditions in both directions, all the edge states disappear for $r<r_c=\frac{1}{\sqrt{2}}$. The red dotted line corresponds with the critical point $r_c=\frac{1}{\sqrt{2}}$ obtained from CC formulation (see section \ref{sec:CC}).}
\label{fig:energy_vs_r_in_real_space}
\end{figure}

}}

\section{System size dependence of $M^*$} 
\label{sec:appenMstar}

\begin{figure}
\centering
\includegraphics[width=\columnwidth]{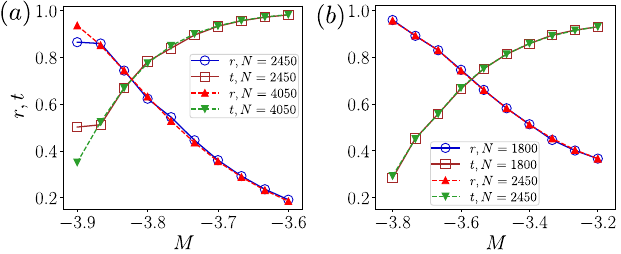}
\caption{{\bf $M^*$ physics:~}(a) $r$ and $ t$ as function of $M$ when $\theta=0$ with $N=2450$ and $N=4050$ with ($\alpha=1, \beta=1,w=5$). (b) $r$ and $ t$ as function of $M$ when $\theta=\frac{\pi}{6}$ with $N=1800$ and $N=2450$ with ($\alpha=1, \beta=1,w=5$). }
\label{fig:parametersS5}
\end{figure}

In \Fig{fig:parametersS5} we show the system size dependence of $r,t$ as a function of $M$ for a fixed $\theta = 0, \frac{\pi}{6}$. We find that the critical point $M^*$ (where $r=t$) is independent of the system size. We further note that near $M \sim -4$, where the bulk gap closes, system size needs to be significantly large to reduce finite size effects.

\bibliography{refkirigami, ref1DCorr}

\begin{thebibliography}{70}%
\makeatletter
\providecommand \@ifxundefined [1]{%
 \@ifx{#1\undefined}
}%
\providecommand \@ifnum [1]{%
 \ifnum #1\expandafter \@firstoftwo
 \else \expandafter \@secondoftwo
 \fi
}%
\providecommand \@ifx [1]{%
 \ifx #1\expandafter \@firstoftwo
 \else \expandafter \@secondoftwo
 \fi
}%
\providecommand \natexlab [1]{#1}%
\providecommand \enquote  [1]{``#1''}%
\providecommand \bibnamefont  [1]{#1}%
\providecommand \bibfnamefont [1]{#1}%
\providecommand \citenamefont [1]{#1}%
\providecommand \href@noop [0]{\@secondoftwo}%
\providecommand \href [0]{\begingroup \@sanitize@url \@href}%
\providecommand \@href[1]{\@@startlink{#1}\@@href}%
\providecommand \@@href[1]{\endgroup#1\@@endlink}%
\providecommand \@sanitize@url [0]{\catcode `\\12\catcode `\$12\catcode
  `\&12\catcode `\#12\catcode `\^12\catcode `\_12\catcode `\%12\relax}%
\providecommand \@@startlink[1]{}%
\providecommand \@@endlink[0]{}%
\providecommand \url  [0]{\begingroup\@sanitize@url \@url }%
\providecommand \@url [1]{\endgroup\@href {#1}{\urlprefix }}%
\providecommand \urlprefix  [0]{URL }%
\providecommand \Eprint [0]{\href }%
\providecommand \doibase [0]{https://doi.org/}%
\providecommand \selectlanguage [0]{\@gobble}%
\providecommand \bibinfo  [0]{\@secondoftwo}%
\providecommand \bibfield  [0]{\@secondoftwo}%
\providecommand \translation [1]{[#1]}%
\providecommand \BibitemOpen [0]{}%
\providecommand \bibitemStop [0]{}%
\providecommand \bibitemNoStop [0]{.\EOS\space}%
\providecommand \EOS [0]{\spacefactor3000\relax}%
\providecommand \BibitemShut  [1]{\csname bibitem#1\endcsname}%
\let\auto@bib@innerbib\@empty
\bibitem [{\citenamefont {Rogers}\ \emph {et~al.}(2010)\citenamefont {Rogers},
  \citenamefont {Someya},\ and\ \citenamefont {Huang}}]{rogers2010materials}%
  \BibitemOpen
  \bibfield  {author} {\bibinfo {author} {\bibfnamefont {J.~A.}\ \bibnamefont
  {Rogers}}, \bibinfo {author} {\bibfnamefont {T.}~\bibnamefont {Someya}},\
  and\ \bibinfo {author} {\bibfnamefont {Y.}~\bibnamefont {Huang}},\ }\bibfield
   {title} {\bibinfo {title} {Materials and mechanics for stretchable
  electronics},\ }\href {https://doi.org/DOI: 10.1126/science.1182383}
  {\bibfield  {journal} {\bibinfo  {journal} {science}\ }\textbf {\bibinfo
  {volume} {327}},\ \bibinfo {pages} {1603} (\bibinfo {year}
  {2010})}\BibitemShut {NoStop}%
\bibitem [{\citenamefont {Santangelo}(2023)}]{santangelo2023making}%
  \BibitemOpen
  \bibfield  {author} {\bibinfo {author} {\bibfnamefont {C.~D.}\ \bibnamefont
  {Santangelo}},\ }\bibfield  {title} {\bibinfo {title} {Making smarter
  materials},\ }\href
  {https://doi.org/https://doi.org/10.1038/s41563-022-01408-w} {\bibfield
  {journal} {\bibinfo  {journal} {Nature Materials}\ }\textbf {\bibinfo
  {volume} {22}},\ \bibinfo {pages} {3} (\bibinfo {year} {2023})}\BibitemShut
  {NoStop}%
\bibitem [{\citenamefont {El~Helou}\ \emph {et~al.}(2022)\citenamefont
  {El~Helou}, \citenamefont {Grossmann}, \citenamefont {Tabor}, \citenamefont
  {Buskohl},\ and\ \citenamefont {Harne}}]{el2022mechanical}%
  \BibitemOpen
  \bibfield  {author} {\bibinfo {author} {\bibfnamefont {C.}~\bibnamefont
  {El~Helou}}, \bibinfo {author} {\bibfnamefont {B.}~\bibnamefont {Grossmann}},
  \bibinfo {author} {\bibfnamefont {C.~E.}\ \bibnamefont {Tabor}}, \bibinfo
  {author} {\bibfnamefont {P.~R.}\ \bibnamefont {Buskohl}},\ and\ \bibinfo
  {author} {\bibfnamefont {R.~L.}\ \bibnamefont {Harne}},\ }\bibfield  {title}
  {\bibinfo {title} {Mechanical integrated circuit materials},\ }\href
  {https://doi.org/https://doi.org/10.1038/s41586-022-05004-5} {\bibfield
  {journal} {\bibinfo  {journal} {Nature}\ }\textbf {\bibinfo {volume} {608}},\
  \bibinfo {pages} {699} (\bibinfo {year} {2022})}\BibitemShut {NoStop}%
\bibitem [{\citenamefont {El~Helou}\ \emph {et~al.}(2021)\citenamefont
  {El~Helou}, \citenamefont {Buskohl}, \citenamefont {Tabor},\ and\
  \citenamefont {Harne}}]{el2021digital}%
  \BibitemOpen
  \bibfield  {author} {\bibinfo {author} {\bibfnamefont {C.}~\bibnamefont
  {El~Helou}}, \bibinfo {author} {\bibfnamefont {P.~R.}\ \bibnamefont
  {Buskohl}}, \bibinfo {author} {\bibfnamefont {C.~E.}\ \bibnamefont {Tabor}},\
  and\ \bibinfo {author} {\bibfnamefont {R.~L.}\ \bibnamefont {Harne}},\
  }\bibfield  {title} {\bibinfo {title} {Digital logic gates in soft,
  conductive mechanical metamaterials},\ }\href
  {https://doi.org/https://doi.org/10.1038/s41467-021-21920-y} {\bibfield
  {journal} {\bibinfo  {journal} {Nature communications}\ }\textbf {\bibinfo
  {volume} {12}},\ \bibinfo {pages} {1633} (\bibinfo {year}
  {2021})}\BibitemShut {NoStop}%
\bibitem [{\citenamefont {Zhai}\ \emph {et~al.}(2021)\citenamefont {Zhai},
  \citenamefont {Wu},\ and\ \citenamefont {Jiang}}]{zhai2021mechanical}%
  \BibitemOpen
  \bibfield  {author} {\bibinfo {author} {\bibfnamefont {Z.}~\bibnamefont
  {Zhai}}, \bibinfo {author} {\bibfnamefont {L.}~\bibnamefont {Wu}},\ and\
  \bibinfo {author} {\bibfnamefont {H.}~\bibnamefont {Jiang}},\ }\bibfield
  {title} {\bibinfo {title} {Mechanical metamaterials based on origami and
  kirigami},\ }\href {https://doi.org/10.1063/5.0051088} {\bibfield  {journal}
  {\bibinfo  {journal} {Applied Physics Reviews}\ }\textbf {\bibinfo {volume}
  {8}} (\bibinfo {year} {2021})}\BibitemShut {NoStop}%
\bibitem [{\citenamefont {Chen}\ \emph {et~al.}(2020)\citenamefont {Chen},
  \citenamefont {Chen}, \citenamefont {Zhang}, \citenamefont {Li},\ and\
  \citenamefont {Li}}]{chen2020kirigami}%
  \BibitemOpen
  \bibfield  {author} {\bibinfo {author} {\bibfnamefont {S.}~\bibnamefont
  {Chen}}, \bibinfo {author} {\bibfnamefont {J.}~\bibnamefont {Chen}}, \bibinfo
  {author} {\bibfnamefont {X.}~\bibnamefont {Zhang}}, \bibinfo {author}
  {\bibfnamefont {Z.-Y.}\ \bibnamefont {Li}},\ and\ \bibinfo {author}
  {\bibfnamefont {J.}~\bibnamefont {Li}},\ }\bibfield  {title} {\bibinfo
  {title} {Kirigami/origami: unfolding the new regime of advanced 3d
  microfabrication/nanofabrication with “folding”},\ }\href
  {https://doi.org/https://doi.org/10.1038/s41377-020-0309-9} {\bibfield
  {journal} {\bibinfo  {journal} {Light: Science \& Applications}\ }\textbf
  {\bibinfo {volume} {9}},\ \bibinfo {pages} {75} (\bibinfo {year}
  {2020})}\BibitemShut {NoStop}%
\bibitem [{\citenamefont {Blees}\ \emph {et~al.}(2015)\citenamefont {Blees},
  \citenamefont {Barnard}, \citenamefont {Rose}, \citenamefont {Roberts},
  \citenamefont {McGill}, \citenamefont {Huang}, \citenamefont {Ruyack},
  \citenamefont {Kevek}, \citenamefont {Kobrin}, \citenamefont {Muller} \emph
  {et~al.}}]{blees2015graphene}%
  \BibitemOpen
  \bibfield  {author} {\bibinfo {author} {\bibfnamefont {M.~K.}\ \bibnamefont
  {Blees}}, \bibinfo {author} {\bibfnamefont {A.~W.}\ \bibnamefont {Barnard}},
  \bibinfo {author} {\bibfnamefont {P.~A.}\ \bibnamefont {Rose}}, \bibinfo
  {author} {\bibfnamefont {S.~P.}\ \bibnamefont {Roberts}}, \bibinfo {author}
  {\bibfnamefont {K.~L.}\ \bibnamefont {McGill}}, \bibinfo {author}
  {\bibfnamefont {P.~Y.}\ \bibnamefont {Huang}}, \bibinfo {author}
  {\bibfnamefont {A.~R.}\ \bibnamefont {Ruyack}}, \bibinfo {author}
  {\bibfnamefont {J.~W.}\ \bibnamefont {Kevek}}, \bibinfo {author}
  {\bibfnamefont {B.}~\bibnamefont {Kobrin}}, \bibinfo {author} {\bibfnamefont
  {D.~A.}\ \bibnamefont {Muller}}, \emph {et~al.},\ }\bibfield  {title}
  {\bibinfo {title} {Graphene kirigami},\ }\href
  {https://doi.org/10.1038/nature14588} {\bibfield  {journal} {\bibinfo
  {journal} {Nature}\ }\textbf {\bibinfo {volume} {524}},\ \bibinfo {pages}
  {204} (\bibinfo {year} {2015})}\BibitemShut {NoStop}%
\bibitem [{\citenamefont {Ouyang}\ \emph {et~al.}(2023)\citenamefont {Ouyang},
  \citenamefont {Gu}, \citenamefont {Gao}, \citenamefont {Hu}, \citenamefont
  {Zhang}, \citenamefont {Ren}, \citenamefont {Li}, \citenamefont {Sun},
  \citenamefont {Chen},\ and\ \citenamefont {Ding}}]{KirigamiPRA2023}%
  \BibitemOpen
  \bibfield  {author} {\bibinfo {author} {\bibfnamefont {H.}~\bibnamefont
  {Ouyang}}, \bibinfo {author} {\bibfnamefont {Y.}~\bibnamefont {Gu}}, \bibinfo
  {author} {\bibfnamefont {Z.}~\bibnamefont {Gao}}, \bibinfo {author}
  {\bibfnamefont {L.}~\bibnamefont {Hu}}, \bibinfo {author} {\bibfnamefont
  {Z.}~\bibnamefont {Zhang}}, \bibinfo {author} {\bibfnamefont
  {J.}~\bibnamefont {Ren}}, \bibinfo {author} {\bibfnamefont {B.}~\bibnamefont
  {Li}}, \bibinfo {author} {\bibfnamefont {J.}~\bibnamefont {Sun}}, \bibinfo
  {author} {\bibfnamefont {Y.}~\bibnamefont {Chen}},\ and\ \bibinfo {author}
  {\bibfnamefont {X.}~\bibnamefont {Ding}},\ }\bibfield  {title} {\bibinfo
  {title} {Kirigami-inspired thermal regulator},\ }\href
  {https://doi.org/10.1103/PhysRevApplied.19.L011001} {\bibfield  {journal}
  {\bibinfo  {journal} {Phys. Rev. Appl.}\ }\textbf {\bibinfo {volume} {19}},\
  \bibinfo {pages} {L011001} (\bibinfo {year} {2023})}\BibitemShut {NoStop}%
\bibitem [{\citenamefont {Liu}\ \emph {et~al.}(2018)\citenamefont {Liu},
  \citenamefont {Du}, \citenamefont {Li}, \citenamefont {Lu}, \citenamefont
  {Li},\ and\ \citenamefont {Fang}}]{liu2018nano}%
  \BibitemOpen
  \bibfield  {author} {\bibinfo {author} {\bibfnamefont {Z.}~\bibnamefont
  {Liu}}, \bibinfo {author} {\bibfnamefont {H.}~\bibnamefont {Du}}, \bibinfo
  {author} {\bibfnamefont {J.}~\bibnamefont {Li}}, \bibinfo {author}
  {\bibfnamefont {L.}~\bibnamefont {Lu}}, \bibinfo {author} {\bibfnamefont
  {Z.-Y.}\ \bibnamefont {Li}},\ and\ \bibinfo {author} {\bibfnamefont {N.~X.}\
  \bibnamefont {Fang}},\ }\bibfield  {title} {\bibinfo {title} {Nano-kirigami
  with giant optical chirality},\ }\href {https://doi.org/DOI:
  10.1126/sciadv.aat4436} {\bibfield  {journal} {\bibinfo  {journal} {Science
  advances}\ }\textbf {\bibinfo {volume} {4}},\ \bibinfo {pages} {eaat4436}
  (\bibinfo {year} {2018})}\BibitemShut {NoStop}%
\bibitem [{\citenamefont {Jin}\ and\ \citenamefont
  {Yang}()}]{LishuaiADvMat2023}%
  \BibitemOpen
  \bibfield  {author} {\bibinfo {author} {\bibfnamefont {L.}~\bibnamefont
  {Jin}}\ and\ \bibinfo {author} {\bibfnamefont {S.}~\bibnamefont {Yang}},\
  }\bibfield  {title} {\bibinfo {title} {Engineering kirigami frameworks toward
  real-world applications},\ }\href
  {https://doi.org/https://doi.org/10.1002/adma.202308560} {\bibfield
  {journal} {\bibinfo  {journal} {Advanced Materials}\ }\textbf {\bibinfo
  {volume} {n/a}},\ \bibinfo {pages} {2308560}}\BibitemShut {NoStop}%
\bibitem [{\citenamefont {Castro}\ \emph {et~al.}(2018)\citenamefont {Castro},
  \citenamefont {Flachi}, \citenamefont {Ribeiro},\ and\ \citenamefont
  {Vitagliano}}]{Castro_PRL_2018}%
  \BibitemOpen
  \bibfield  {author} {\bibinfo {author} {\bibfnamefont {E.~V.}\ \bibnamefont
  {Castro}}, \bibinfo {author} {\bibfnamefont {A.}~\bibnamefont {Flachi}},
  \bibinfo {author} {\bibfnamefont {P.}~\bibnamefont {Ribeiro}},\ and\ \bibinfo
  {author} {\bibfnamefont {V.}~\bibnamefont {Vitagliano}},\ }\bibfield  {title}
  {\bibinfo {title} {Symmetry breaking and lattice kirigami},\ }\href
  {https://doi.org/10.1103/PhysRevLett.121.221601} {\bibfield  {journal}
  {\bibinfo  {journal} {Phys. Rev. Lett.}\ }\textbf {\bibinfo {volume} {121}},\
  \bibinfo {pages} {221601} (\bibinfo {year} {2018})}\BibitemShut {NoStop}%
\bibitem [{\citenamefont {Flachi}\ and\ \citenamefont
  {Vitagliano}(2019)}]{Flachi_PRD_2019}%
  \BibitemOpen
  \bibfield  {author} {\bibinfo {author} {\bibfnamefont {A.}~\bibnamefont
  {Flachi}}\ and\ \bibinfo {author} {\bibfnamefont {V.}~\bibnamefont
  {Vitagliano}},\ }\bibfield  {title} {\bibinfo {title} {Symmetry breaking and
  lattice kirigami: Finite temperature effects},\ }\href
  {https://doi.org/10.1103/PhysRevD.99.125010} {\bibfield  {journal} {\bibinfo
  {journal} {Phys. Rev. D}\ }\textbf {\bibinfo {volume} {99}},\ \bibinfo
  {pages} {125010} (\bibinfo {year} {2019})}\BibitemShut {NoStop}%
\bibitem [{\citenamefont {Imaki}\ and\ \citenamefont
  {Yamamoto}(2019)}]{Imaki_PRD_2019}%
  \BibitemOpen
  \bibfield  {author} {\bibinfo {author} {\bibfnamefont {S.}~\bibnamefont
  {Imaki}}\ and\ \bibinfo {author} {\bibfnamefont {A.}~\bibnamefont
  {Yamamoto}},\ }\bibfield  {title} {\bibinfo {title} {Lattice field theory
  with torsion},\ }\href {https://doi.org/10.1103/PhysRevD.100.054509}
  {\bibfield  {journal} {\bibinfo  {journal} {Phys. Rev. D}\ }\textbf {\bibinfo
  {volume} {100}},\ \bibinfo {pages} {054509} (\bibinfo {year}
  {2019})}\BibitemShut {NoStop}%
\bibitem [{\citenamefont {Grosso}\ and\ \citenamefont
  {Mele}(2015)}]{GrossoPRL2015}%
  \BibitemOpen
  \bibfield  {author} {\bibinfo {author} {\bibfnamefont {B.~F.}\ \bibnamefont
  {Grosso}}\ and\ \bibinfo {author} {\bibfnamefont {E.~J.}\ \bibnamefont
  {Mele}},\ }\bibfield  {title} {\bibinfo {title} {Bending rules in graphene
  kirigami},\ }\href {https://doi.org/10.1103/PhysRevLett.115.195501}
  {\bibfield  {journal} {\bibinfo  {journal} {Phys. Rev. Lett.}\ }\textbf
  {\bibinfo {volume} {115}},\ \bibinfo {pages} {195501} (\bibinfo {year}
  {2015})}\BibitemShut {NoStop}%
\bibitem [{\citenamefont {Hanakata}\ \emph {et~al.}(2018)\citenamefont
  {Hanakata}, \citenamefont {Cubuk}, \citenamefont {Campbell},\ and\
  \citenamefont {Park}}]{Hanakata_PRL_2018}%
  \BibitemOpen
  \bibfield  {author} {\bibinfo {author} {\bibfnamefont {P.~Z.}\ \bibnamefont
  {Hanakata}}, \bibinfo {author} {\bibfnamefont {E.~D.}\ \bibnamefont {Cubuk}},
  \bibinfo {author} {\bibfnamefont {D.~K.}\ \bibnamefont {Campbell}},\ and\
  \bibinfo {author} {\bibfnamefont {H.~S.}\ \bibnamefont {Park}},\ }\bibfield
  {title} {\bibinfo {title} {Accelerated search and design of stretchable
  graphene kirigami using machine learning},\ }\href
  {https://doi.org/10.1103/PhysRevLett.121.255304} {\bibfield  {journal}
  {\bibinfo  {journal} {Phys. Rev. Lett.}\ }\textbf {\bibinfo {volume} {121}},\
  \bibinfo {pages} {255304} (\bibinfo {year} {2018})}\BibitemShut {NoStop}%
\bibitem [{\citenamefont {Bahamon}\ \emph {et~al.}(2016)\citenamefont
  {Bahamon}, \citenamefont {Qi}, \citenamefont {Park}, \citenamefont
  {Pereira},\ and\ \citenamefont {Campbell}}]{Campbell_kirigami_2016}%
  \BibitemOpen
  \bibfield  {author} {\bibinfo {author} {\bibfnamefont {D.~A.}\ \bibnamefont
  {Bahamon}}, \bibinfo {author} {\bibfnamefont {Z.}~\bibnamefont {Qi}},
  \bibinfo {author} {\bibfnamefont {H.~S.}\ \bibnamefont {Park}}, \bibinfo
  {author} {\bibfnamefont {V.~M.}\ \bibnamefont {Pereira}},\ and\ \bibinfo
  {author} {\bibfnamefont {D.~K.}\ \bibnamefont {Campbell}},\ }\bibfield
  {title} {\bibinfo {title} {Graphene kirigami as a platform for stretchable
  and tunable quantum dot arrays},\ }\href
  {https://doi.org/10.1103/PhysRevB.93.235408} {\bibfield  {journal} {\bibinfo
  {journal} {Phys. Rev. B}\ }\textbf {\bibinfo {volume} {93}},\ \bibinfo
  {pages} {235408} (\bibinfo {year} {2016})}\BibitemShut {NoStop}%
\bibitem [{\citenamefont {Qi}\ \emph {et~al.}(2014)\citenamefont {Qi},
  \citenamefont {Campbell},\ and\ \citenamefont
  {Park}}]{Campbell_kirigami_2014}%
  \BibitemOpen
  \bibfield  {author} {\bibinfo {author} {\bibfnamefont {Z.}~\bibnamefont
  {Qi}}, \bibinfo {author} {\bibfnamefont {D.~K.}\ \bibnamefont {Campbell}},\
  and\ \bibinfo {author} {\bibfnamefont {H.~S.}\ \bibnamefont {Park}},\
  }\bibfield  {title} {\bibinfo {title} {Atomistic simulations of
  tension-induced large deformation and stretchability in graphene kirigami},\
  }\href {https://doi.org/10.1103/PhysRevB.90.245437} {\bibfield  {journal}
  {\bibinfo  {journal} {Phys. Rev. B}\ }\textbf {\bibinfo {volume} {90}},\
  \bibinfo {pages} {245437} (\bibinfo {year} {2014})}\BibitemShut {NoStop}%
\bibitem [{\citenamefont {Liu}\ \emph {et~al.}(2022{\natexlab{a}})\citenamefont
  {Liu}, \citenamefont {Choi},\ and\ \citenamefont {Mahadevan}}]{Liu_PRR_2022}%
  \BibitemOpen
  \bibfield  {author} {\bibinfo {author} {\bibfnamefont {L.}~\bibnamefont
  {Liu}}, \bibinfo {author} {\bibfnamefont {G.~P.~T.}\ \bibnamefont {Choi}},\
  and\ \bibinfo {author} {\bibfnamefont {L.}~\bibnamefont {Mahadevan}},\
  }\bibfield  {title} {\bibinfo {title} {Quasicrystal kirigami},\ }\href
  {https://doi.org/10.1103/PhysRevResearch.4.033114} {\bibfield  {journal}
  {\bibinfo  {journal} {Phys. Rev. Res.}\ }\textbf {\bibinfo {volume} {4}},\
  \bibinfo {pages} {033114} (\bibinfo {year} {2022}{\natexlab{a}})}\BibitemShut
  {NoStop}%
\bibitem [{\citenamefont {Chen}\ \emph {et~al.}(2016)\citenamefont {Chen},
  \citenamefont {Liu}, \citenamefont {Evans}, \citenamefont {Paulose},
  \citenamefont {Cohen}, \citenamefont {Vitelli},\ and\ \citenamefont
  {Santangelo}}]{Chen2016Topological}%
  \BibitemOpen
  \bibfield  {author} {\bibinfo {author} {\bibfnamefont {B.~G.-g.}\
  \bibnamefont {Chen}}, \bibinfo {author} {\bibfnamefont {B.}~\bibnamefont
  {Liu}}, \bibinfo {author} {\bibfnamefont {A.~A.}\ \bibnamefont {Evans}},
  \bibinfo {author} {\bibfnamefont {J.}~\bibnamefont {Paulose}}, \bibinfo
  {author} {\bibfnamefont {I.}~\bibnamefont {Cohen}}, \bibinfo {author}
  {\bibfnamefont {V.}~\bibnamefont {Vitelli}},\ and\ \bibinfo {author}
  {\bibfnamefont {C.~D.}\ \bibnamefont {Santangelo}},\ }\bibfield  {title}
  {\bibinfo {title} {Topological mechanics of origami and kirigami},\ }\href
  {https://doi.org/10.1103/PhysRevLett.116.135501} {\bibfield  {journal}
  {\bibinfo  {journal} {Phys. Rev. Lett.}\ }\textbf {\bibinfo {volume} {116}},\
  \bibinfo {pages} {135501} (\bibinfo {year} {2016})}\BibitemShut {NoStop}%
\bibitem [{\citenamefont {Mortazavi}\ \emph {et~al.}(2017)\citenamefont
  {Mortazavi}, \citenamefont {Lherbier}, \citenamefont {Fan}, \citenamefont
  {Harju}, \citenamefont {Rabczuk},\ and\ \citenamefont
  {Charlier}}]{mortazavi2017thermal}%
  \BibitemOpen
  \bibfield  {author} {\bibinfo {author} {\bibfnamefont {B.}~\bibnamefont
  {Mortazavi}}, \bibinfo {author} {\bibfnamefont {A.}~\bibnamefont {Lherbier}},
  \bibinfo {author} {\bibfnamefont {Z.}~\bibnamefont {Fan}}, \bibinfo {author}
  {\bibfnamefont {A.}~\bibnamefont {Harju}}, \bibinfo {author} {\bibfnamefont
  {T.}~\bibnamefont {Rabczuk}},\ and\ \bibinfo {author} {\bibfnamefont {J.-C.}\
  \bibnamefont {Charlier}},\ }\bibfield  {title} {\bibinfo {title} {Thermal and
  electronic transport characteristics of highly stretchable graphene
  kirigami},\ }\href {https://doi.org/10.1039/C7NR05231F} {\bibfield  {journal}
  {\bibinfo  {journal} {Nanoscale}\ }\textbf {\bibinfo {volume} {9}},\ \bibinfo
  {pages} {16329} (\bibinfo {year} {2017})}\BibitemShut {NoStop}%
\bibitem [{\citenamefont {Zheng}\ \emph {et~al.}(2022)\citenamefont {Zheng},
  \citenamefont {Niloy}, \citenamefont {Celli}, \citenamefont {Tobasco},\ and\
  \citenamefont {Plucinsky}}]{zheng2022continuum}%
  \BibitemOpen
  \bibfield  {author} {\bibinfo {author} {\bibfnamefont {Y.}~\bibnamefont
  {Zheng}}, \bibinfo {author} {\bibfnamefont {I.}~\bibnamefont {Niloy}},
  \bibinfo {author} {\bibfnamefont {P.}~\bibnamefont {Celli}}, \bibinfo
  {author} {\bibfnamefont {I.}~\bibnamefont {Tobasco}},\ and\ \bibinfo {author}
  {\bibfnamefont {P.}~\bibnamefont {Plucinsky}},\ }\bibfield  {title} {\bibinfo
  {title} {Continuum field theory for the deformations of planar kirigami},\
  }\href {https://doi.org/10.1103/PhysRevLett.128.208003} {\bibfield  {journal}
  {\bibinfo  {journal} {Physical Review Letters}\ }\textbf {\bibinfo {volume}
  {128}},\ \bibinfo {pages} {208003} (\bibinfo {year} {2022})}\BibitemShut
  {NoStop}%
\bibitem [{\citenamefont {Ludwig}(2016)}]{Ludwig_PS_2015}%
  \BibitemOpen
  \bibfield  {author} {\bibinfo {author} {\bibfnamefont {A.~W.~W.}\
  \bibnamefont {Ludwig}},\ }\bibfield  {title} {\bibinfo {title} {Topological
  phases: classification of topological insulators and superconductors of
  non-interacting fermions, and beyond},\ }\href
  {http://stacks.iop.org/1402-4896/2016/i=T168/a=014001} {\bibfield  {journal}
  {\bibinfo  {journal} {Physica Scripta}\ }\textbf {\bibinfo {volume} {2016}},\
  \bibinfo {pages} {014001} (\bibinfo {year} {2016})}\BibitemShut {NoStop}%
\bibitem [{\citenamefont {Chiu}\ \emph {et~al.}(2016)\citenamefont {Chiu},
  \citenamefont {Teo}, \citenamefont {Schnyder},\ and\ \citenamefont
  {Ryu}}]{Chiu_RMP_2016}%
  \BibitemOpen
  \bibfield  {author} {\bibinfo {author} {\bibfnamefont {C.-K.}\ \bibnamefont
  {Chiu}}, \bibinfo {author} {\bibfnamefont {J.~C.~Y.}\ \bibnamefont {Teo}},
  \bibinfo {author} {\bibfnamefont {A.~P.}\ \bibnamefont {Schnyder}},\ and\
  \bibinfo {author} {\bibfnamefont {S.}~\bibnamefont {Ryu}},\ }\bibfield
  {title} {\bibinfo {title} {Classification of topological quantum matter with
  symmetries},\ }\href {https://doi.org/10.1103/RevModPhys.88.035005}
  {\bibfield  {journal} {\bibinfo  {journal} {Rev. Mod. Phys.}\ }\textbf
  {\bibinfo {volume} {88}},\ \bibinfo {pages} {035005} (\bibinfo {year}
  {2016})}\BibitemShut {NoStop}%
\bibitem [{\citenamefont {Hasan}\ and\ \citenamefont
  {Kane}(2010)}]{Hasan_RMP_2010}%
  \BibitemOpen
  \bibfield  {author} {\bibinfo {author} {\bibfnamefont {M.~Z.}\ \bibnamefont
  {Hasan}}\ and\ \bibinfo {author} {\bibfnamefont {C.~L.}\ \bibnamefont
  {Kane}},\ }\bibfield  {title} {\bibinfo {title} {\textit{Colloquium} :
  Topological insulators},\ }\href {https://doi.org/10.1103/RevModPhys.82.3045}
  {\bibfield  {journal} {\bibinfo  {journal} {Rev. Mod. Phys.}\ }\textbf
  {\bibinfo {volume} {82}},\ \bibinfo {pages} {3045} (\bibinfo {year}
  {2010})}\BibitemShut {NoStop}%
\bibitem [{\citenamefont {Qi}\ and\ \citenamefont {Zhang}(2011)}]{Qi_RMP_2011}%
  \BibitemOpen
  \bibfield  {author} {\bibinfo {author} {\bibfnamefont {X.-L.}\ \bibnamefont
  {Qi}}\ and\ \bibinfo {author} {\bibfnamefont {S.-C.}\ \bibnamefont {Zhang}},\
  }\bibfield  {title} {\bibinfo {title} {Topological insulators and
  superconductors},\ }\href {https://doi.org/10.1103/RevModPhys.83.1057}
  {\bibfield  {journal} {\bibinfo  {journal} {Rev. Mod. Phys.}\ }\textbf
  {\bibinfo {volume} {83}},\ \bibinfo {pages} {1057} (\bibinfo {year}
  {2011})}\BibitemShut {NoStop}%
\bibitem [{\citenamefont {Asb{\'o}th}\ \emph {et~al.}(2016)\citenamefont
  {Asb{\'o}th}, \citenamefont {Oroszl{\'a}ny},\ and\ \citenamefont
  {P{\'a}lyi}}]{asboth2016short}%
  \BibitemOpen
  \bibfield  {author} {\bibinfo {author} {\bibfnamefont {J.~K.}\ \bibnamefont
  {Asb{\'o}th}}, \bibinfo {author} {\bibfnamefont {L.}~\bibnamefont
  {Oroszl{\'a}ny}},\ and\ \bibinfo {author} {\bibfnamefont {A.}~\bibnamefont
  {P{\'a}lyi}},\ }\href@noop {} {\emph {\bibinfo {title} {A short course on
  topological insulators}}}\ (\bibinfo  {publisher} {Springer Nature},\
  \bibinfo {year} {2016})\BibitemShut {NoStop}%
\bibitem [{\citenamefont {Chalker}\ and\ \citenamefont
  {Coddington}(1988)}]{chalker1988percolation}%
  \BibitemOpen
  \bibfield  {author} {\bibinfo {author} {\bibfnamefont {J.}~\bibnamefont
  {Chalker}}\ and\ \bibinfo {author} {\bibfnamefont {P.}~\bibnamefont
  {Coddington}},\ }\bibfield  {title} {\bibinfo {title} {Percolation, quantum
  tunnelling and the integer hall effect},\ }\href
  {https://doi.org/10.1088/0022-3719/21/14/008} {\bibfield  {journal} {\bibinfo
   {journal} {Journal of Physics C: Solid State Physics}\ }\textbf {\bibinfo
  {volume} {21}},\ \bibinfo {pages} {2665} (\bibinfo {year}
  {1988})}\BibitemShut {NoStop}%
\bibitem [{\citenamefont {Ho}\ and\ \citenamefont
  {Chalker}(1996)}]{Chalker1996networkmodels}%
  \BibitemOpen
  \bibfield  {author} {\bibinfo {author} {\bibfnamefont {C.-M.}\ \bibnamefont
  {Ho}}\ and\ \bibinfo {author} {\bibfnamefont {J.~T.}\ \bibnamefont
  {Chalker}},\ }\bibfield  {title} {\bibinfo {title} {Models for the integer
  quantum hall effect: The network model, the dirac equation, and a
  tight-binding hamiltonian},\ }\href
  {https://doi.org/10.1103/PhysRevB.54.8708} {\bibfield  {journal} {\bibinfo
  {journal} {Phys. Rev. B}\ }\textbf {\bibinfo {volume} {54}},\ \bibinfo
  {pages} {8708} (\bibinfo {year} {1996})}\BibitemShut {NoStop}%
\bibitem [{\citenamefont {Kramer}\ \emph {et~al.}(2005)\citenamefont {Kramer},
  \citenamefont {Ohtsuki},\ and\ \citenamefont {Kettemann}}]{kramer2005random}%
  \BibitemOpen
  \bibfield  {author} {\bibinfo {author} {\bibfnamefont {B.}~\bibnamefont
  {Kramer}}, \bibinfo {author} {\bibfnamefont {T.}~\bibnamefont {Ohtsuki}},\
  and\ \bibinfo {author} {\bibfnamefont {S.}~\bibnamefont {Kettemann}},\
  }\bibfield  {title} {\bibinfo {title} {Random network models and quantum
  phase transitions in two dimensions},\ }\href
  {https://doi.org/https://doi.org/10.1016/j.physrep.2005.07.001} {\bibfield
  {journal} {\bibinfo  {journal} {Physics reports}\ }\textbf {\bibinfo {volume}
  {417}},\ \bibinfo {pages} {211} (\bibinfo {year} {2005})}\BibitemShut
  {NoStop}%
\bibitem [{\citenamefont {Lee}\ \emph {et~al.}(1993)\citenamefont {Lee},
  \citenamefont {Wang},\ and\ \citenamefont {Kivelson}}]{Lee_PRL_1993}%
  \BibitemOpen
  \bibfield  {author} {\bibinfo {author} {\bibfnamefont {D.-H.}\ \bibnamefont
  {Lee}}, \bibinfo {author} {\bibfnamefont {Z.}~\bibnamefont {Wang}},\ and\
  \bibinfo {author} {\bibfnamefont {S.}~\bibnamefont {Kivelson}},\ }\bibfield
  {title} {\bibinfo {title} {Quantum percolation and plateau transitions in the
  quantum hall effect},\ }\href {https://doi.org/10.1103/PhysRevLett.70.4130}
  {\bibfield  {journal} {\bibinfo  {journal} {Phys. Rev. Lett.}\ }\textbf
  {\bibinfo {volume} {70}},\ \bibinfo {pages} {4130} (\bibinfo {year}
  {1993})}\BibitemShut {NoStop}%
\bibitem [{\citenamefont {Snyman}\ \emph {et~al.}(2008)\citenamefont {Snyman},
  \citenamefont {Tworzyd\l{}o},\ and\ \citenamefont
  {Beenakker}}]{Snyman_PRB_2008}%
  \BibitemOpen
  \bibfield  {author} {\bibinfo {author} {\bibfnamefont {I.}~\bibnamefont
  {Snyman}}, \bibinfo {author} {\bibfnamefont {J.}~\bibnamefont
  {Tworzyd\l{}o}},\ and\ \bibinfo {author} {\bibfnamefont {C.~W.~J.}\
  \bibnamefont {Beenakker}},\ }\bibfield  {title} {\bibinfo {title}
  {Calculation of the conductance of a graphene sheet using the
  chalker-coddington network model},\ }\href
  {https://doi.org/10.1103/PhysRevB.78.045118} {\bibfield  {journal} {\bibinfo
  {journal} {Phys. Rev. B}\ }\textbf {\bibinfo {volume} {78}},\ \bibinfo
  {pages} {045118} (\bibinfo {year} {2008})}\BibitemShut {NoStop}%
\bibitem [{\citenamefont {Cho}\ and\ \citenamefont
  {Fisher}(1997)}]{Cho_PRB_1997}%
  \BibitemOpen
  \bibfield  {author} {\bibinfo {author} {\bibfnamefont {S.}~\bibnamefont
  {Cho}}\ and\ \bibinfo {author} {\bibfnamefont {M.~P.~A.}\ \bibnamefont
  {Fisher}},\ }\bibfield  {title} {\bibinfo {title} {Criticality in the
  two-dimensional random-bond ising model},\ }\href
  {https://doi.org/10.1103/PhysRevB.55.1025} {\bibfield  {journal} {\bibinfo
  {journal} {Phys. Rev. B}\ }\textbf {\bibinfo {volume} {55}},\ \bibinfo
  {pages} {1025} (\bibinfo {year} {1997})}\BibitemShut {NoStop}%
\bibitem [{\citenamefont {Charles}\ \emph {et~al.}(2020)\citenamefont
  {Charles}, \citenamefont {Gruzberg}, \citenamefont {Kl\"umper}, \citenamefont
  {Nuding},\ and\ \citenamefont {Sedrakyan}}]{GruzbergPRBkagome}%
  \BibitemOpen
  \bibfield  {author} {\bibinfo {author} {\bibfnamefont {N.}~\bibnamefont
  {Charles}}, \bibinfo {author} {\bibfnamefont {I.~A.}\ \bibnamefont
  {Gruzberg}}, \bibinfo {author} {\bibfnamefont {A.}~\bibnamefont {Kl\"umper}},
  \bibinfo {author} {\bibfnamefont {W.}~\bibnamefont {Nuding}},\ and\ \bibinfo
  {author} {\bibfnamefont {A.}~\bibnamefont {Sedrakyan}},\ }\bibfield  {title}
  {\bibinfo {title} {Critical behavior at the integer quantum hall transition
  in a network model on the kagome lattice},\ }\href
  {https://doi.org/10.1103/PhysRevB.102.121304} {\bibfield  {journal} {\bibinfo
   {journal} {Phys. Rev. B}\ }\textbf {\bibinfo {volume} {102}},\ \bibinfo
  {pages} {121304} (\bibinfo {year} {2020})}\BibitemShut {NoStop}%
\bibitem [{\citenamefont {Potter}\ \emph {et~al.}(2020)\citenamefont {Potter},
  \citenamefont {Chalker},\ and\ \citenamefont
  {Gurarie}}]{ChalkerPRL2020Floquet}%
  \BibitemOpen
  \bibfield  {author} {\bibinfo {author} {\bibfnamefont {A.~C.}\ \bibnamefont
  {Potter}}, \bibinfo {author} {\bibfnamefont {J.~T.}\ \bibnamefont
  {Chalker}},\ and\ \bibinfo {author} {\bibfnamefont {V.}~\bibnamefont
  {Gurarie}},\ }\bibfield  {title} {\bibinfo {title} {Quantum hall network
  models as floquet topological insulators},\ }\href
  {https://doi.org/10.1103/PhysRevLett.125.086601} {\bibfield  {journal}
  {\bibinfo  {journal} {Phys. Rev. Lett.}\ }\textbf {\bibinfo {volume} {125}},\
  \bibinfo {pages} {086601} (\bibinfo {year} {2020})}\BibitemShut {NoStop}%
\bibitem [{\citenamefont {Obuse}\ \emph {et~al.}(2012)\citenamefont {Obuse},
  \citenamefont {Gruzberg},\ and\ \citenamefont
  {Evers}}]{GruzbergPRLFinite-Size}%
  \BibitemOpen
  \bibfield  {author} {\bibinfo {author} {\bibfnamefont {H.}~\bibnamefont
  {Obuse}}, \bibinfo {author} {\bibfnamefont {I.~A.}\ \bibnamefont
  {Gruzberg}},\ and\ \bibinfo {author} {\bibfnamefont {F.}~\bibnamefont
  {Evers}},\ }\bibfield  {title} {\bibinfo {title} {Finite-size effects and
  irrelevant corrections to scaling near the integer quantum hall transition},\
  }\href {https://doi.org/10.1103/PhysRevLett.109.206804} {\bibfield  {journal}
  {\bibinfo  {journal} {Phys. Rev. Lett.}\ }\textbf {\bibinfo {volume} {109}},\
  \bibinfo {pages} {206804} (\bibinfo {year} {2012})}\BibitemShut {NoStop}%
\bibitem [{\citenamefont {Hu}\ \emph {et~al.}(2015)\citenamefont {Hu},
  \citenamefont {Pillay}, \citenamefont {Wu}, \citenamefont {Pasek},
  \citenamefont {Shum},\ and\ \citenamefont {Chong}}]{PhysRevX2015Measurement}%
  \BibitemOpen
  \bibfield  {author} {\bibinfo {author} {\bibfnamefont {W.}~\bibnamefont
  {Hu}}, \bibinfo {author} {\bibfnamefont {J.~C.}\ \bibnamefont {Pillay}},
  \bibinfo {author} {\bibfnamefont {K.}~\bibnamefont {Wu}}, \bibinfo {author}
  {\bibfnamefont {M.}~\bibnamefont {Pasek}}, \bibinfo {author} {\bibfnamefont
  {P.~P.}\ \bibnamefont {Shum}},\ and\ \bibinfo {author} {\bibfnamefont
  {Y.~D.}\ \bibnamefont {Chong}},\ }\bibfield  {title} {\bibinfo {title}
  {Measurement of a topological edge invariant in a microwave network},\ }\href
  {https://doi.org/10.1103/PhysRevX.5.011012} {\bibfield  {journal} {\bibinfo
  {journal} {Phys. Rev. X}\ }\textbf {\bibinfo {volume} {5}},\ \bibinfo {pages}
  {011012} (\bibinfo {year} {2015})}\BibitemShut {NoStop}%
\bibitem [{\citenamefont {Chalker}\ \emph {et~al.}(2001)\citenamefont
  {Chalker}, \citenamefont {Read}, \citenamefont {Kagalovsky}, \citenamefont
  {Horovitz}, \citenamefont {Avishai},\ and\ \citenamefont
  {Ludwig}}]{Chalker2001Thermalnetwork}%
  \BibitemOpen
  \bibfield  {author} {\bibinfo {author} {\bibfnamefont {J.~T.}\ \bibnamefont
  {Chalker}}, \bibinfo {author} {\bibfnamefont {N.}~\bibnamefont {Read}},
  \bibinfo {author} {\bibfnamefont {V.}~\bibnamefont {Kagalovsky}}, \bibinfo
  {author} {\bibfnamefont {B.}~\bibnamefont {Horovitz}}, \bibinfo {author}
  {\bibfnamefont {Y.}~\bibnamefont {Avishai}},\ and\ \bibinfo {author}
  {\bibfnamefont {A.~W.~W.}\ \bibnamefont {Ludwig}},\ }\bibfield  {title}
  {\bibinfo {title} {Thermal metal in network models of a disordered
  two-dimensional superconductor},\ }\href
  {https://doi.org/10.1103/PhysRevB.65.012506} {\bibfield  {journal} {\bibinfo
  {journal} {Phys. Rev. B}\ }\textbf {\bibinfo {volume} {65}},\ \bibinfo
  {pages} {012506} (\bibinfo {year} {2001})}\BibitemShut {NoStop}%
\bibitem [{\citenamefont {Pasek}\ and\ \citenamefont
  {Chong}(2014)}]{Pasek2014Networkmodels}%
  \BibitemOpen
  \bibfield  {author} {\bibinfo {author} {\bibfnamefont {M.}~\bibnamefont
  {Pasek}}\ and\ \bibinfo {author} {\bibfnamefont {Y.~D.}\ \bibnamefont
  {Chong}},\ }\bibfield  {title} {\bibinfo {title} {Network models of photonic
  floquet topological insulators},\ }\href
  {https://doi.org/10.1103/PhysRevB.89.075113} {\bibfield  {journal} {\bibinfo
  {journal} {Phys. Rev. B}\ }\textbf {\bibinfo {volume} {89}},\ \bibinfo
  {pages} {075113} (\bibinfo {year} {2014})}\BibitemShut {NoStop}%
\bibitem [{\citenamefont {De~Beule}\ \emph {et~al.}(2021)\citenamefont
  {De~Beule}, \citenamefont {Dominguez},\ and\ \citenamefont
  {Recher}}]{Bilayergraphenenetwork2021}%
  \BibitemOpen
  \bibfield  {author} {\bibinfo {author} {\bibfnamefont {C.}~\bibnamefont
  {De~Beule}}, \bibinfo {author} {\bibfnamefont {F.}~\bibnamefont
  {Dominguez}},\ and\ \bibinfo {author} {\bibfnamefont {P.}~\bibnamefont
  {Recher}},\ }\bibfield  {title} {\bibinfo {title} {Network model and
  four-terminal transport in minimally twisted bilayer graphene},\ }\href
  {https://doi.org/10.1103/PhysRevB.104.195410} {\bibfield  {journal} {\bibinfo
   {journal} {Phys. Rev. B}\ }\textbf {\bibinfo {volume} {104}},\ \bibinfo
  {pages} {195410} (\bibinfo {year} {2021})}\BibitemShut {NoStop}%
\bibitem [{\citenamefont {Bernevig}\ \emph {et~al.}(2006)\citenamefont
  {Bernevig}, \citenamefont {Hughes},\ and\ \citenamefont
  {Zhang}}]{bernevig2006quantum}%
  \BibitemOpen
  \bibfield  {author} {\bibinfo {author} {\bibfnamefont {B.~A.}\ \bibnamefont
  {Bernevig}}, \bibinfo {author} {\bibfnamefont {T.~L.}\ \bibnamefont
  {Hughes}},\ and\ \bibinfo {author} {\bibfnamefont {S.-C.}\ \bibnamefont
  {Zhang}},\ }\bibfield  {title} {\bibinfo {title} {Quantum spin hall effect
  and topological phase transition in hgte quantum wells},\ }\href
  {https://doi.org/10.1126/science.1133734} {\bibfield  {journal} {\bibinfo
  {journal} {science}\ }\textbf {\bibinfo {volume} {314}},\ \bibinfo {pages}
  {1757} (\bibinfo {year} {2006})}\BibitemShut {NoStop}%
\bibitem [{\citenamefont {Tao}\ \emph {et~al.}(2023)\citenamefont {Tao},
  \citenamefont {Khosravi}, \citenamefont {Deshpande},\ and\ \citenamefont
  {Li}}]{tao2023Engineering}%
  \BibitemOpen
  \bibfield  {author} {\bibinfo {author} {\bibfnamefont {J.}~\bibnamefont
  {Tao}}, \bibinfo {author} {\bibfnamefont {H.}~\bibnamefont {Khosravi}},
  \bibinfo {author} {\bibfnamefont {V.}~\bibnamefont {Deshpande}},\ and\
  \bibinfo {author} {\bibfnamefont {S.}~\bibnamefont {Li}},\ }\bibfield
  {title} {\bibinfo {title} {Engineering by cuts: How kirigami principle
  enables unique mechanical properties and functionalities},\ }\href
  {https://doi.org/10.1002/advs.202204733} {\bibfield  {journal} {\bibinfo
  {journal} {Advanced Science}\ }\textbf {\bibinfo {volume} {10}},\ \bibinfo
  {pages} {2204733} (\bibinfo {year} {2023})}\BibitemShut {NoStop}%
\bibitem [{\citenamefont {Rafsanjani}\ and\ \citenamefont
  {Bertoldi}(2017)}]{rafsanjani2017buckling}%
  \BibitemOpen
  \bibfield  {author} {\bibinfo {author} {\bibfnamefont {A.}~\bibnamefont
  {Rafsanjani}}\ and\ \bibinfo {author} {\bibfnamefont {K.}~\bibnamefont
  {Bertoldi}},\ }\bibfield  {title} {\bibinfo {title} {Buckling-induced
  kirigami},\ }\href {https://link.aps.org/doi/10.1103/PhysRevLett.118.084301}
  {\bibfield  {journal} {\bibinfo  {journal} {Physical review letters}\
  }\textbf {\bibinfo {volume} {118}},\ \bibinfo {pages} {084301} (\bibinfo
  {year} {2017})}\BibitemShut {NoStop}%
\bibitem [{\citenamefont {Lee}\ \emph {et~al.}(2020)\citenamefont {Lee},
  \citenamefont {Hsieh}, \citenamefont {Yong},\ and\ \citenamefont
  {Nam}}]{lee2020multiaxially}%
  \BibitemOpen
  \bibfield  {author} {\bibinfo {author} {\bibfnamefont {H.~C.}\ \bibnamefont
  {Lee}}, \bibinfo {author} {\bibfnamefont {E.~Y.}\ \bibnamefont {Hsieh}},
  \bibinfo {author} {\bibfnamefont {K.}~\bibnamefont {Yong}},\ and\ \bibinfo
  {author} {\bibfnamefont {S.}~\bibnamefont {Nam}},\ }\bibfield  {title}
  {\bibinfo {title} {Multiaxially-stretchable kirigami-patterned mesh design
  for graphene sensor devices},\ }\href
  {https://doi.org/10.1007/s12274-020-2662-7} {\bibfield  {journal} {\bibinfo
  {journal} {Nano Research}\ }\textbf {\bibinfo {volume} {13}},\ \bibinfo
  {pages} {1406} (\bibinfo {year} {2020})}\BibitemShut {NoStop}%
\bibitem [{\citenamefont {Grima}\ and\ \citenamefont
  {Evans}(2000)}]{grima2000auxetic}%
  \BibitemOpen
  \bibfield  {author} {\bibinfo {author} {\bibfnamefont {J.~N.}\ \bibnamefont
  {Grima}}\ and\ \bibinfo {author} {\bibfnamefont {K.~E.}\ \bibnamefont
  {Evans}},\ }\bibfield  {title} {\bibinfo {title} {Auxetic behavior from
  rotating squares},\ }\href {https://doi.org/10.1023/A:1006781224002}
  {\bibfield  {journal} {\bibinfo  {journal} {Journal of materials science
  letters}\ }\textbf {\bibinfo {volume} {19}},\ \bibinfo {pages} {1563}
  (\bibinfo {year} {2000})}\BibitemShut {NoStop}%
\bibitem [{\citenamefont {Shuaibu}\ \emph {et~al.}(2024)\citenamefont
  {Shuaibu}, \citenamefont {Deng}, \citenamefont {Xu}, \citenamefont {Ade-Oke},
  \citenamefont {Aliyu},\ and\ \citenamefont {Momoh}}]{shuaibu2024advancing}%
  \BibitemOpen
  \bibfield  {author} {\bibinfo {author} {\bibfnamefont {A.~S.}\ \bibnamefont
  {Shuaibu}}, \bibinfo {author} {\bibfnamefont {J.}~\bibnamefont {Deng}},
  \bibinfo {author} {\bibfnamefont {C.}~\bibnamefont {Xu}}, \bibinfo {author}
  {\bibfnamefont {V.~P.}\ \bibnamefont {Ade-Oke}}, \bibinfo {author}
  {\bibfnamefont {A.}~\bibnamefont {Aliyu}},\ and\ \bibinfo {author}
  {\bibfnamefont {D.}~\bibnamefont {Momoh}},\ }\bibfield  {title} {\bibinfo
  {title} {Advancing auxetic materials: Emerging development and innovative
  applications},\ }\href {https://doi.org/10.1515/rams-2024-0021} {\bibfield
  {journal} {\bibinfo  {journal} {Reviews on Advanced Materials Science}\
  }\textbf {\bibinfo {volume} {63}},\ \bibinfo {pages} {20240021} (\bibinfo
  {year} {2024})}\BibitemShut {NoStop}%
\bibitem [{num()}]{numcomment}%
  \BibitemOpen
  \href@noop {} {}\bibinfo {note} {Given $a=1$, and a typical $\xi=10a$, the
  edge state localization length; assuming $L= 40a$ and if the network contains
  $40 \times 40$ Kirigami blocks ($L_B=40$) performing wavepacket dynamics for
  the full network requires a simulation on a Hamiltonian of size $N_T \sim
  10^6$. On the other hand the multiscale analysis reduces this to a
  calculation of just $\sim L^2_B$ even on a system with no translational
  symmetry.}\BibitemShut {Stop}%
\bibitem [{\citenamefont {Hatano}\ and\ \citenamefont
  {Nelson}(1996)}]{HatanoPhysRevLett}%
  \BibitemOpen
  \bibfield  {author} {\bibinfo {author} {\bibfnamefont {N.}~\bibnamefont
  {Hatano}}\ and\ \bibinfo {author} {\bibfnamefont {D.~R.}\ \bibnamefont
  {Nelson}},\ }\bibfield  {title} {\bibinfo {title} {Localization transitions
  in non-hermitian quantum mechanics},\ }\href
  {https://doi.org/10.1103/PhysRevLett.77.570} {\bibfield  {journal} {\bibinfo
  {journal} {Phys. Rev. Lett.}\ }\textbf {\bibinfo {volume} {77}},\ \bibinfo
  {pages} {570} (\bibinfo {year} {1996})}\BibitemShut {NoStop}%
\bibitem [{\citenamefont {Hatano}\ and\ \citenamefont
  {Nelson}(1997)}]{HatanoPhysRevB}%
  \BibitemOpen
  \bibfield  {author} {\bibinfo {author} {\bibfnamefont {N.}~\bibnamefont
  {Hatano}}\ and\ \bibinfo {author} {\bibfnamefont {D.~R.}\ \bibnamefont
  {Nelson}},\ }\bibfield  {title} {\bibinfo {title} {Vortex pinning and
  non-hermitian quantum mechanics},\ }\href
  {https://doi.org/10.1103/PhysRevB.56.8651} {\bibfield  {journal} {\bibinfo
  {journal} {Phys. Rev. B}\ }\textbf {\bibinfo {volume} {56}},\ \bibinfo
  {pages} {8651} (\bibinfo {year} {1997})}\BibitemShut {NoStop}%
\bibitem [{\citenamefont {Lee}\ \emph {et~al.}(2019)\citenamefont {Lee},
  \citenamefont {Ahn}, \citenamefont {Zhou},\ and\ \citenamefont
  {Vishwanath}}]{VishwanathPhysRevLett}%
  \BibitemOpen
  \bibfield  {author} {\bibinfo {author} {\bibfnamefont {J.~Y.}\ \bibnamefont
  {Lee}}, \bibinfo {author} {\bibfnamefont {J.}~\bibnamefont {Ahn}}, \bibinfo
  {author} {\bibfnamefont {H.}~\bibnamefont {Zhou}},\ and\ \bibinfo {author}
  {\bibfnamefont {A.}~\bibnamefont {Vishwanath}},\ }\bibfield  {title}
  {\bibinfo {title} {Topological correspondence between hermitian and
  non-hermitian systems: Anomalous dynamics},\ }\href
  {https://doi.org/10.1103/PhysRevLett.123.206404} {\bibfield  {journal}
  {\bibinfo  {journal} {Phys. Rev. Lett.}\ }\textbf {\bibinfo {volume} {123}},\
  \bibinfo {pages} {206404} (\bibinfo {year} {2019})}\BibitemShut {NoStop}%
\bibitem [{\citenamefont {Lieu}(2018)}]{SimonLieuPhysRevB}%
  \BibitemOpen
  \bibfield  {author} {\bibinfo {author} {\bibfnamefont {S.}~\bibnamefont
  {Lieu}},\ }\bibfield  {title} {\bibinfo {title} {Topological phases in the
  non-hermitian su-schrieffer-heeger model},\ }\href
  {https://doi.org/10.1103/PhysRevB.97.045106} {\bibfield  {journal} {\bibinfo
  {journal} {Phys. Rev. B}\ }\textbf {\bibinfo {volume} {97}},\ \bibinfo
  {pages} {045106} (\bibinfo {year} {2018})}\BibitemShut {NoStop}%
\bibitem [{\citenamefont {Shen}\ \emph {et~al.}(2018)\citenamefont {Shen},
  \citenamefont {Zhen},\ and\ \citenamefont {Fu}}]{PhysRevLett.120.146402}%
  \BibitemOpen
  \bibfield  {author} {\bibinfo {author} {\bibfnamefont {H.}~\bibnamefont
  {Shen}}, \bibinfo {author} {\bibfnamefont {B.}~\bibnamefont {Zhen}},\ and\
  \bibinfo {author} {\bibfnamefont {L.}~\bibnamefont {Fu}},\ }\bibfield
  {title} {\bibinfo {title} {Topological band theory for non-hermitian
  hamiltonians},\ }\href {https://doi.org/10.1103/PhysRevLett.120.146402}
  {\bibfield  {journal} {\bibinfo  {journal} {Phys. Rev. Lett.}\ }\textbf
  {\bibinfo {volume} {120}},\ \bibinfo {pages} {146402} (\bibinfo {year}
  {2018})}\BibitemShut {NoStop}%
\bibitem [{\citenamefont {Schindler}\ \emph {et~al.}(2023)\citenamefont
  {Schindler}, \citenamefont {Gu}, \citenamefont {Lian},\ and\ \citenamefont
  {Kawabata}}]{PRXQuantum.4.030315}%
  \BibitemOpen
  \bibfield  {author} {\bibinfo {author} {\bibfnamefont {F.}~\bibnamefont
  {Schindler}}, \bibinfo {author} {\bibfnamefont {K.}~\bibnamefont {Gu}},
  \bibinfo {author} {\bibfnamefont {B.}~\bibnamefont {Lian}},\ and\ \bibinfo
  {author} {\bibfnamefont {K.}~\bibnamefont {Kawabata}},\ }\bibfield  {title}
  {\bibinfo {title} {Hermitian bulk -- non-hermitian boundary correspondence},\
  }\href {https://doi.org/10.1103/PRXQuantum.4.030315} {\bibfield  {journal}
  {\bibinfo  {journal} {PRX Quantum}\ }\textbf {\bibinfo {volume} {4}},\
  \bibinfo {pages} {030315} (\bibinfo {year} {2023})}\BibitemShut {NoStop}%
\bibitem [{\citenamefont {Dey}\ \emph {et~al.}(2024)\citenamefont {Dey},
  \citenamefont {Shahriar}, \citenamefont {Anan}, \citenamefont {Malakar},
  \citenamefont {Rahman},\ and\ \citenamefont {Chowdhury}}]{dey2024enhancing}%
  \BibitemOpen
  \bibfield  {author} {\bibinfo {author} {\bibfnamefont {K.}~\bibnamefont
  {Dey}}, \bibinfo {author} {\bibfnamefont {S.}~\bibnamefont {Shahriar}},
  \bibinfo {author} {\bibfnamefont {M.}~\bibnamefont {Anan}}, \bibinfo {author}
  {\bibfnamefont {P.}~\bibnamefont {Malakar}}, \bibinfo {author} {\bibfnamefont
  {M.}~\bibnamefont {Rahman}},\ and\ \bibinfo {author} {\bibfnamefont
  {M.}~\bibnamefont {Chowdhury}},\ }\bibfield  {title} {\bibinfo {title}
  {Enhancing the stretchability of two-dimensional materials through kirigami:
  a molecular dynamics study on tungsten disulfide},\ }\href
  {https://doi.org/10.1039/d4ra04814h} {\bibfield  {journal} {\bibinfo
  {journal} {RSC advances}\ }\textbf {\bibinfo {volume} {14}},\ \bibinfo
  {pages} {24483} (\bibinfo {year} {2024})}\BibitemShut {NoStop}%
\bibitem [{\citenamefont {Han}\ \emph {et~al.}(2017)\citenamefont {Han},
  \citenamefont {Scarpa},\ and\ \citenamefont {Allan}}]{han2017super}%
  \BibitemOpen
  \bibfield  {author} {\bibinfo {author} {\bibfnamefont {T.}~\bibnamefont
  {Han}}, \bibinfo {author} {\bibfnamefont {F.}~\bibnamefont {Scarpa}},\ and\
  \bibinfo {author} {\bibfnamefont {N.~L.}\ \bibnamefont {Allan}},\ }\bibfield
  {title} {\bibinfo {title} {Super stretchable hexagonal boron nitride
  kirigami},\ }\href {https://doi.org/10.1016/j.tsf.2017.03.059} {\bibfield
  {journal} {\bibinfo  {journal} {Thin Solid Films}\ }\textbf {\bibinfo
  {volume} {632}},\ \bibinfo {pages} {35} (\bibinfo {year} {2017})}\BibitemShut
  {NoStop}%
\bibitem [{\citenamefont {Kumar}\ \emph {et~al.}(2020)\citenamefont {Kumar},
  \citenamefont {Mishra},\ and\ \citenamefont
  {Mahata}}]{kumar2020manipulation}%
  \BibitemOpen
  \bibfield  {author} {\bibinfo {author} {\bibfnamefont {S.}~\bibnamefont
  {Kumar}}, \bibinfo {author} {\bibfnamefont {T.}~\bibnamefont {Mishra}},\ and\
  \bibinfo {author} {\bibfnamefont {A.}~\bibnamefont {Mahata}},\ }\bibfield
  {title} {\bibinfo {title} {Manipulation of mechanical properties of monolayer
  molybdenum disulfide: Kirigami and hetero-structure based approach},\ }\href
  {https://doi.org/10.1016/j.matchemphys.2020.123280} {\bibfield  {journal}
  {\bibinfo  {journal} {Materials Chemistry and Physics}\ }\textbf {\bibinfo
  {volume} {252}},\ \bibinfo {pages} {123280} (\bibinfo {year}
  {2020})}\BibitemShut {NoStop}%
\bibitem [{\citenamefont {Zhu}\ \emph {et~al.}(2024)\citenamefont {Zhu},
  \citenamefont {Wang}, \citenamefont {Zhang}, \citenamefont {Zhao},
  \citenamefont {Yu},\ and\ \citenamefont {Zhang}}]{zhu2024mechanical}%
  \BibitemOpen
  \bibfield  {author} {\bibinfo {author} {\bibfnamefont {P.}~\bibnamefont
  {Zhu}}, \bibinfo {author} {\bibfnamefont {S.}~\bibnamefont {Wang}}, \bibinfo
  {author} {\bibfnamefont {X.}~\bibnamefont {Zhang}}, \bibinfo {author}
  {\bibfnamefont {J.}~\bibnamefont {Zhao}}, \bibinfo {author} {\bibfnamefont
  {W.}~\bibnamefont {Yu}},\ and\ \bibinfo {author} {\bibfnamefont
  {H.}~\bibnamefont {Zhang}},\ }\bibfield  {title} {\bibinfo {title}
  {Mechanical properties of diamane kirigami under tensile deformation},\
  }\href {https://doi.org/10.1007/s11051-024-06004-4} {\bibfield  {journal}
  {\bibinfo  {journal} {Journal of Nanoparticle Research}\ }\textbf {\bibinfo
  {volume} {26}},\ \bibinfo {pages} {91} (\bibinfo {year} {2024})}\BibitemShut
  {NoStop}%
\bibitem [{\citenamefont {Pezo}\ \emph {et~al.}(2021)\citenamefont {Pezo},
  \citenamefont {Focassio}, \citenamefont {Schleder}, \citenamefont {Costa},
  \citenamefont {Lewenkopf},\ and\ \citenamefont {Fazzio}}]{Pezo_PRM_2021}%
  \BibitemOpen
  \bibfield  {author} {\bibinfo {author} {\bibfnamefont {A.}~\bibnamefont
  {Pezo}}, \bibinfo {author} {\bibfnamefont {B.}~\bibnamefont {Focassio}},
  \bibinfo {author} {\bibfnamefont {G.~R.}\ \bibnamefont {Schleder}}, \bibinfo
  {author} {\bibfnamefont {M.}~\bibnamefont {Costa}}, \bibinfo {author}
  {\bibfnamefont {C.}~\bibnamefont {Lewenkopf}},\ and\ \bibinfo {author}
  {\bibfnamefont {A.}~\bibnamefont {Fazzio}},\ }\bibfield  {title} {\bibinfo
  {title} {Disorder effects of vacancies on the electronic transport properties
  of realistic topological insulator nanoribbons: The case of bismuthene},\
  }\href {https://doi.org/10.1103/PhysRevMaterials.5.014204} {\bibfield
  {journal} {\bibinfo  {journal} {Phys. Rev. Mater.}\ }\textbf {\bibinfo
  {volume} {5}},\ \bibinfo {pages} {014204} (\bibinfo {year}
  {2021})}\BibitemShut {NoStop}%
\bibitem [{\citenamefont {Reis}\ \emph {et~al.}(2017)\citenamefont {Reis},
  \citenamefont {Li}, \citenamefont {Dudy}, \citenamefont {Bauernfeind},
  \citenamefont {Glass}, \citenamefont {Hanke}, \citenamefont {Thomale},
  \citenamefont {Sch{\"a}fer},\ and\ \citenamefont
  {Claessen}}]{reis2017bismuthene}%
  \BibitemOpen
  \bibfield  {author} {\bibinfo {author} {\bibfnamefont {F.}~\bibnamefont
  {Reis}}, \bibinfo {author} {\bibfnamefont {G.}~\bibnamefont {Li}}, \bibinfo
  {author} {\bibfnamefont {L.}~\bibnamefont {Dudy}}, \bibinfo {author}
  {\bibfnamefont {M.}~\bibnamefont {Bauernfeind}}, \bibinfo {author}
  {\bibfnamefont {S.}~\bibnamefont {Glass}}, \bibinfo {author} {\bibfnamefont
  {W.}~\bibnamefont {Hanke}}, \bibinfo {author} {\bibfnamefont
  {R.}~\bibnamefont {Thomale}}, \bibinfo {author} {\bibfnamefont
  {J.}~\bibnamefont {Sch{\"a}fer}},\ and\ \bibinfo {author} {\bibfnamefont
  {R.}~\bibnamefont {Claessen}},\ }\bibfield  {title} {\bibinfo {title}
  {Bismuthene on a sic substrate: A candidate for a high-temperature quantum
  spin hall material},\ }\href
  {https://www.science.org/doi/abs/10.1126/science.aai8142} {\bibfield
  {journal} {\bibinfo  {journal} {Science}\ }\textbf {\bibinfo {volume}
  {357}},\ \bibinfo {pages} {287} (\bibinfo {year} {2017})}\BibitemShut
  {NoStop}%
\bibitem [{\citenamefont {Focassio}\ \emph {et~al.}(2021)\citenamefont
  {Focassio}, \citenamefont {Schleder}, \citenamefont {Costa}, \citenamefont
  {Fazzio},\ and\ \citenamefont {Lewenkopf}}]{focassio2021structural}%
  \BibitemOpen
  \bibfield  {author} {\bibinfo {author} {\bibfnamefont {B.}~\bibnamefont
  {Focassio}}, \bibinfo {author} {\bibfnamefont {G.~R.}\ \bibnamefont
  {Schleder}}, \bibinfo {author} {\bibfnamefont {M.}~\bibnamefont {Costa}},
  \bibinfo {author} {\bibfnamefont {A.}~\bibnamefont {Fazzio}},\ and\ \bibinfo
  {author} {\bibfnamefont {C.}~\bibnamefont {Lewenkopf}},\ }\bibfield  {title}
  {\bibinfo {title} {Structural and electronic properties of realistic
  two-dimensional amorphous topological insulators},\ }\href
  {https://doi.org/DOI 10.1088/2053-1583/abdb97} {\bibfield  {journal}
  {\bibinfo  {journal} {2D Materials}\ }\textbf {\bibinfo {volume} {8}},\
  \bibinfo {pages} {025032} (\bibinfo {year} {2021})}\BibitemShut {NoStop}%
\bibitem [{\citenamefont {Kozlov}\ \emph {et~al.}(2014)\citenamefont {Kozlov},
  \citenamefont {Kvon}, \citenamefont {Olshanetsky}, \citenamefont {Mikhailov},
  \citenamefont {Dvoretsky},\ and\ \citenamefont {Weiss}}]{Kozlov_PRL_2014}%
  \BibitemOpen
  \bibfield  {author} {\bibinfo {author} {\bibfnamefont {D.~A.}\ \bibnamefont
  {Kozlov}}, \bibinfo {author} {\bibfnamefont {Z.~D.}\ \bibnamefont {Kvon}},
  \bibinfo {author} {\bibfnamefont {E.~B.}\ \bibnamefont {Olshanetsky}},
  \bibinfo {author} {\bibfnamefont {N.~N.}\ \bibnamefont {Mikhailov}}, \bibinfo
  {author} {\bibfnamefont {S.~A.}\ \bibnamefont {Dvoretsky}},\ and\ \bibinfo
  {author} {\bibfnamefont {D.}~\bibnamefont {Weiss}},\ }\bibfield  {title}
  {\bibinfo {title} {Transport properties of a 3d topological insulator based
  on a strained high-mobility hgte film},\ }\href
  {https://doi.org/10.1103/PhysRevLett.112.196801} {\bibfield  {journal}
  {\bibinfo  {journal} {Phys. Rev. Lett.}\ }\textbf {\bibinfo {volume} {112}},\
  \bibinfo {pages} {196801} (\bibinfo {year} {2014})}\BibitemShut {NoStop}%
\bibitem [{\citenamefont {Olshanetsky}\ \emph {et~al.}(2015)\citenamefont
  {Olshanetsky}, \citenamefont {Kvon}, \citenamefont {Gusev}, \citenamefont
  {Levin}, \citenamefont {Raichev}, \citenamefont {Mikhailov},\ and\
  \citenamefont {Dvoretsky}}]{Olsh_PRL_2015}%
  \BibitemOpen
  \bibfield  {author} {\bibinfo {author} {\bibfnamefont {E.~B.}\ \bibnamefont
  {Olshanetsky}}, \bibinfo {author} {\bibfnamefont {Z.~D.}\ \bibnamefont
  {Kvon}}, \bibinfo {author} {\bibfnamefont {G.~M.}\ \bibnamefont {Gusev}},
  \bibinfo {author} {\bibfnamefont {A.~D.}\ \bibnamefont {Levin}}, \bibinfo
  {author} {\bibfnamefont {O.~E.}\ \bibnamefont {Raichev}}, \bibinfo {author}
  {\bibfnamefont {N.~N.}\ \bibnamefont {Mikhailov}},\ and\ \bibinfo {author}
  {\bibfnamefont {S.~A.}\ \bibnamefont {Dvoretsky}},\ }\bibfield  {title}
  {\bibinfo {title} {Persistence of a two-dimensional topological insulator
  state in wide hgte quantum wells},\ }\href
  {https://doi.org/10.1103/PhysRevLett.114.126802} {\bibfield  {journal}
  {\bibinfo  {journal} {Phys. Rev. Lett.}\ }\textbf {\bibinfo {volume} {114}},\
  \bibinfo {pages} {126802} (\bibinfo {year} {2015})}\BibitemShut {NoStop}%
\bibitem [{\citenamefont {Kvon}\ \emph {et~al.}(2020)\citenamefont {Kvon},
  \citenamefont {Kozlov}, \citenamefont {Olshanetsky}, \citenamefont {Gusev},
  \citenamefont {Mikhailov},\ and\ \citenamefont
  {Dvoretsky}}]{kvon2020topological}%
  \BibitemOpen
  \bibfield  {author} {\bibinfo {author} {\bibfnamefont {Z.~D.}\ \bibnamefont
  {Kvon}}, \bibinfo {author} {\bibfnamefont {D.~A.}\ \bibnamefont {Kozlov}},
  \bibinfo {author} {\bibfnamefont {E.}~\bibnamefont {Olshanetsky}}, \bibinfo
  {author} {\bibfnamefont {G.}~\bibnamefont {Gusev}}, \bibinfo {author}
  {\bibfnamefont {N.~N.}\ \bibnamefont {Mikhailov}},\ and\ \bibinfo {author}
  {\bibfnamefont {S.}~\bibnamefont {Dvoretsky}},\ }\bibfield  {title} {\bibinfo
  {title} {Topological insulators based on hgte},\ }\href@noop {} {\bibfield
  {journal} {\bibinfo  {journal} {Physics-Uspekhi}\ }\textbf {\bibinfo {volume}
  {63}},\ \bibinfo {pages} {629} (\bibinfo {year} {2020})}\BibitemShut
  {NoStop}%
\bibitem [{\citenamefont {Liu}\ \emph {et~al.}(2022{\natexlab{b}})\citenamefont
  {Liu}, \citenamefont {Li}, \citenamefont {Wang}, \citenamefont {Wei},
  \citenamefont {Zhu}, \citenamefont {Sun}, \citenamefont {Lin},\ and\
  \citenamefont {Xu}}]{liu2022robust}%
  \BibitemOpen
  \bibfield  {author} {\bibinfo {author} {\bibfnamefont {H.}~\bibnamefont
  {Liu}}, \bibinfo {author} {\bibfnamefont {H.}~\bibnamefont {Li}}, \bibinfo
  {author} {\bibfnamefont {Z.}~\bibnamefont {Wang}}, \bibinfo {author}
  {\bibfnamefont {X.}~\bibnamefont {Wei}}, \bibinfo {author} {\bibfnamefont
  {H.}~\bibnamefont {Zhu}}, \bibinfo {author} {\bibfnamefont {M.}~\bibnamefont
  {Sun}}, \bibinfo {author} {\bibfnamefont {Y.}~\bibnamefont {Lin}},\ and\
  \bibinfo {author} {\bibfnamefont {L.}~\bibnamefont {Xu}},\ }\bibfield
  {title} {\bibinfo {title} {Robust and multifunctional kirigami electronics
  with a tough and permeable aramid nanofiber framework},\ }\href
  {https://doi.org/10.1002/adma.202207350} {\bibfield  {journal} {\bibinfo
  {journal} {Advanced Materials}\ }\textbf {\bibinfo {volume} {34}},\ \bibinfo
  {pages} {2207350} (\bibinfo {year} {2022}{\natexlab{b}})}\BibitemShut
  {NoStop}%
\bibitem [{\citenamefont {Tang}\ \emph {et~al.}(2015)\citenamefont {Tang},
  \citenamefont {Lin}, \citenamefont {Han}, \citenamefont {Qiu}, \citenamefont
  {Yang},\ and\ \citenamefont {Yin}}]{tang2015design}%
  \BibitemOpen
  \bibfield  {author} {\bibinfo {author} {\bibfnamefont {Y.}~\bibnamefont
  {Tang}}, \bibinfo {author} {\bibfnamefont {G.}~\bibnamefont {Lin}}, \bibinfo
  {author} {\bibfnamefont {L.}~\bibnamefont {Han}}, \bibinfo {author}
  {\bibfnamefont {S.}~\bibnamefont {Qiu}}, \bibinfo {author} {\bibfnamefont
  {S.}~\bibnamefont {Yang}},\ and\ \bibinfo {author} {\bibfnamefont
  {J.}~\bibnamefont {Yin}},\ }\bibfield  {title} {\bibinfo {title} {Design of
  hierarchically cut hinges for highly stretchable and reconfigurable
  metamaterials with enhanced strength.},\ }\href
  {https://doi.org/10.1002/adma.201502559} {\bibfield  {journal} {\bibinfo
  {journal} {Advanced Materials (Deerfield Beach, Fla.)}\ }\textbf {\bibinfo
  {volume} {27}},\ \bibinfo {pages} {7181} (\bibinfo {year}
  {2015})}\BibitemShut {NoStop}%
\bibitem [{\citenamefont {Kundu}\ and\ \citenamefont
  {Kundu}(2024)}]{Kundu_PRB_2024}%
  \BibitemOpen
  \bibfield  {author} {\bibinfo {author} {\bibfnamefont {R.}~\bibnamefont
  {Kundu}}\ and\ \bibinfo {author} {\bibfnamefont {A.}~\bibnamefont {Kundu}},\
  }\bibfield  {title} {\bibinfo {title} {Josephson junction of minimally
  twisted bilayer graphene},\ }\href
  {https://doi.org/10.1103/PhysRevB.110.085422} {\bibfield  {journal} {\bibinfo
   {journal} {Phys. Rev. B}\ }\textbf {\bibinfo {volume} {110}},\ \bibinfo
  {pages} {085422} (\bibinfo {year} {2024})}\BibitemShut {NoStop}%
\bibitem [{\citenamefont {Yoo}\ \emph {et~al.}(2019)\citenamefont {Yoo},
  \citenamefont {Engelke}, \citenamefont {Carr}, \citenamefont {Fang},
  \citenamefont {Zhang}, \citenamefont {Cazeaux}, \citenamefont {Sung},
  \citenamefont {Hovden}, \citenamefont {Tsen}, \citenamefont {Taniguchi} \emph
  {et~al.}}]{yoo2019atomic}%
  \BibitemOpen
  \bibfield  {author} {\bibinfo {author} {\bibfnamefont {H.}~\bibnamefont
  {Yoo}}, \bibinfo {author} {\bibfnamefont {R.}~\bibnamefont {Engelke}},
  \bibinfo {author} {\bibfnamefont {S.}~\bibnamefont {Carr}}, \bibinfo {author}
  {\bibfnamefont {S.}~\bibnamefont {Fang}}, \bibinfo {author} {\bibfnamefont
  {K.}~\bibnamefont {Zhang}}, \bibinfo {author} {\bibfnamefont
  {P.}~\bibnamefont {Cazeaux}}, \bibinfo {author} {\bibfnamefont {S.~H.}\
  \bibnamefont {Sung}}, \bibinfo {author} {\bibfnamefont {R.}~\bibnamefont
  {Hovden}}, \bibinfo {author} {\bibfnamefont {A.~W.}\ \bibnamefont {Tsen}},
  \bibinfo {author} {\bibfnamefont {T.}~\bibnamefont {Taniguchi}}, \emph
  {et~al.},\ }\bibfield  {title} {\bibinfo {title} {Atomic and electronic
  reconstruction at the van der waals interface in twisted bilayer graphene},\
  }\href {https://doi.org/10.1038/s41563-019-0346-z} {\bibfield  {journal}
  {\bibinfo  {journal} {Nature materials}\ }\textbf {\bibinfo {volume} {18}},\
  \bibinfo {pages} {448} (\bibinfo {year} {2019})}\BibitemShut {NoStop}%
\bibitem [{\citenamefont {Huang}\ \emph {et~al.}(2018)\citenamefont {Huang},
  \citenamefont {Kim}, \citenamefont {Efimkin}, \citenamefont {Lovorn},
  \citenamefont {Taniguchi}, \citenamefont {Watanabe}, \citenamefont
  {MacDonald}, \citenamefont {Tutuc},\ and\ \citenamefont
  {LeRoy}}]{Huang_PRL_2018}%
  \BibitemOpen
  \bibfield  {author} {\bibinfo {author} {\bibfnamefont {S.}~\bibnamefont
  {Huang}}, \bibinfo {author} {\bibfnamefont {K.}~\bibnamefont {Kim}}, \bibinfo
  {author} {\bibfnamefont {D.~K.}\ \bibnamefont {Efimkin}}, \bibinfo {author}
  {\bibfnamefont {T.}~\bibnamefont {Lovorn}}, \bibinfo {author} {\bibfnamefont
  {T.}~\bibnamefont {Taniguchi}}, \bibinfo {author} {\bibfnamefont
  {K.}~\bibnamefont {Watanabe}}, \bibinfo {author} {\bibfnamefont {A.~H.}\
  \bibnamefont {MacDonald}}, \bibinfo {author} {\bibfnamefont {E.}~\bibnamefont
  {Tutuc}},\ and\ \bibinfo {author} {\bibfnamefont {B.~J.}\ \bibnamefont
  {LeRoy}},\ }\bibfield  {title} {\bibinfo {title} {Topologically protected
  helical states in minimally twisted bilayer graphene},\ }\href
  {https://doi.org/10.1103/PhysRevLett.121.037702} {\bibfield  {journal}
  {\bibinfo  {journal} {Phys. Rev. Lett.}\ }\textbf {\bibinfo {volume} {121}},\
  \bibinfo {pages} {037702} (\bibinfo {year} {2018})}\BibitemShut {NoStop}%
\bibitem [{\citenamefont {De~Beule}\ \emph {et~al.}(2020)\citenamefont
  {De~Beule}, \citenamefont {Dominguez},\ and\ \citenamefont
  {Recher}}]{BeuleRPL2020}%
  \BibitemOpen
  \bibfield  {author} {\bibinfo {author} {\bibfnamefont {C.}~\bibnamefont
  {De~Beule}}, \bibinfo {author} {\bibfnamefont {F.}~\bibnamefont
  {Dominguez}},\ and\ \bibinfo {author} {\bibfnamefont {P.}~\bibnamefont
  {Recher}},\ }\bibfield  {title} {\bibinfo {title} {Aharonov-bohm oscillations
  in minimally twisted bilayer graphene},\ }\href
  {https://doi.org/10.1103/PhysRevLett.125.096402} {\bibfield  {journal}
  {\bibinfo  {journal} {Phys. Rev. Lett.}\ }\textbf {\bibinfo {volume} {125}},\
  \bibinfo {pages} {096402} (\bibinfo {year} {2020})}\BibitemShut {NoStop}%
\bibitem [{\citenamefont {Vakhtel}\ \emph {et~al.}(2022)\citenamefont
  {Vakhtel}, \citenamefont {Oriekhov},\ and\ \citenamefont
  {Beenakker}}]{VakhtelPRB2022}%
  \BibitemOpen
  \bibfield  {author} {\bibinfo {author} {\bibfnamefont {T.}~\bibnamefont
  {Vakhtel}}, \bibinfo {author} {\bibfnamefont {D.~O.}\ \bibnamefont
  {Oriekhov}},\ and\ \bibinfo {author} {\bibfnamefont {C.~W.~J.}\ \bibnamefont
  {Beenakker}},\ }\bibfield  {title} {\bibinfo {title} {Bloch oscillations in
  the magnetoconductance of twisted bilayer graphene},\ }\href
  {https://doi.org/10.1103/PhysRevB.105.L241408} {\bibfield  {journal}
  {\bibinfo  {journal} {Phys. Rev. B}\ }\textbf {\bibinfo {volume} {105}},\
  \bibinfo {pages} {L241408} (\bibinfo {year} {2022})}\BibitemShut {NoStop}%
\bibitem [{\citenamefont {An}\ \emph {et~al.}(2020)\citenamefont {An},
  \citenamefont {Domel}, \citenamefont {Zhou}, \citenamefont {Rafsanjani},\
  and\ \citenamefont {Bertoldi}}]{an2020programmable}%
  \BibitemOpen
  \bibfield  {author} {\bibinfo {author} {\bibfnamefont {N.}~\bibnamefont
  {An}}, \bibinfo {author} {\bibfnamefont {A.~G.}\ \bibnamefont {Domel}},
  \bibinfo {author} {\bibfnamefont {J.}~\bibnamefont {Zhou}}, \bibinfo {author}
  {\bibfnamefont {A.}~\bibnamefont {Rafsanjani}},\ and\ \bibinfo {author}
  {\bibfnamefont {K.}~\bibnamefont {Bertoldi}},\ }\bibfield  {title} {\bibinfo
  {title} {Programmable hierarchical kirigami},\ }\href
  {https://doi.org/10.1002/adfm.201906711} {\bibfield  {journal} {\bibinfo
  {journal} {Advanced Functional Materials}\ }\textbf {\bibinfo {volume}
  {30}},\ \bibinfo {pages} {1906711} (\bibinfo {year} {2020})}\BibitemShut
  {NoStop}%
\end{thebibliography}%

\end{document}